\documentclass[preprint,12pt]{elsarticle}

\usepackage{amssymb}
\usepackage{amsmath}

\usepackage{lineno}

\usepackage{textcomp}
\usepackage{color}
\usepackage{lmodern} 

\usepackage{makecell}

\journal{Journal of Computational Physics}

\begin{document}

\begin{frontmatter}



\title{BO-PBK: A comprehensive solver for dispersion relations of obliquely propagating waves 
in magnetized multi-species plasma with anisotropic loss-cone drift product-bi-kappa distribution}


\author[label1]{Wei Bai} 
\ead{baiwei12@mail.ustc.edu.cn, baiweiphys@gmail.com}

\affiliation[label1]{organization={College of Electrical and Power Engineering, Taiyuan University of Technology},
            city={Taiyuan},
            postcode={030024}, 
            country={China}}
            
\author[label2,label3]{Huasheng Xie} 
\ead{xiehuasheng@enn.cn, huashengxie@gmail.com}

\affiliation[label2]{organization={Hebei Key Laboratory of Compact Fusion},
            city={Langfang},
            postcode={065001}, 
            country={China}}
\affiliation[label3]{organization={ENN Science and Technology Development Co., Ltd.},
            city={Langfang},
            postcode={065001}, 
            country={China}}

\begin{abstract}
We present BO-PBK (BO-Product-Bi-Kappa), a new solver for kinetic dispersion relations of obliquely propagating waves in magnetized plasmas with complex velocity distributions. It reformulates the linearized Vlasov–Maxwell system into a compact eigenvalue problem, enabling direct computation of multiple wave branches and unstable modes without iterative initial-value searches. Key innovations include a unified framework supporting product-bi-kappa, kappa-Maxwellian, bi-Maxwellian, and hybrid distributions with multi-component and loss-cone features; a concise rational-form eigenvalue formulation; and a 2-3 times reduction in matrix dimensions compared to the BO-KM solver, with improved efficiency at larger kappa indices. Benchmark tests confirm accurate reproduction of standard kinetic results and efficient resolution of waves and instabilities. BO-PBK thus provides a computationally efficient tool for wave and stability analysis in space and laboratory plasmas.
\end{abstract}




\end{frontmatter}


\section{Introduction}
\label{sec1}
In laboratory and space plasmas, superthermal particles often follow non-Maxwellian distributions, which are well described by the family of Kappa distributions \cite{Lazar_Poedts2010,Lazar2012b,Maksimovic2021}. These distributions exhibit Maxwellian-like behavior at low energies and a power-law decay at high energies, yielding a high-energy tail substantially broader than that of an exponential decay \cite{Lazar2023,Nicolaou2020}. The modified plasma dispersion function for kappa distributions, introduced by Summers and Thorne \cite{Summers1991,Summers1994}, is essential for analyzing kinetic plasma instabilities, especially in systems with complex distributions such as the bi-kappa (BK) \cite{Summers1991}, product-bi-kappa (PBK) \cite{Shrauner_Feldman_1977,Summers1991,bai2024}, and subtracted-kappa (SK)\cite{Summers2025} distributions. Mace and Hellberg \cite{Mace1995} further generalized this dispersion function by allowing the $\kappa$ index to take any positive real value greater than $3/2$. For solar wind ion and electron data from IMP detection\cite{Shrauner_Feldman_1977}, the PBK distribution may provide a better or at least comparable fit than bi-Maxwellian (BM) or BK distributions \cite{Lazar2012b}. By decoupling the parallel and perpendicular dynamics of plasma particles respect to the magnetic field \cite{Lazar_Poedts2010}, the PBK model enables flexible representation of temperature anisotropy, with independent definitions of 
parallel temperature $T_\parallel$, perpendicular temperature $T_\perp$, and corresponding parameters $\kappa_\parallel$ and $\kappa_\perp$. In the limit $\kappa_\perp \to \infty$, the PBK reduces to the kappa-Maxwellian (KM) distribution form \cite{Hellberg2002,Cattaert2007,Singhal2006,bai2025}; when both $\kappa_\parallel, \kappa_\perp \to \infty$, it converges to the BM distribution \cite{bai2024}.

Numerous solvers based on linearized Vlasov-Maxwell equations have been developed for Maxwellian and non-Maxwellian plasmas \cite{Maksimovic2021} . These include solvers for BM (e.g., WHAMP \cite{1982wham}, PLADAWAN \cite{Vinas2000}, PLUME \cite{Klein2015}, NHDS \cite{Verscharen2018_NDHS}, PDRK \cite{xie2016}, BO \cite{xie2019}),  BK (e.g., DHSARK \cite{Astfalk2015}, DIS-K \cite{López_Shaaban_Lazar_2021}, parallel drift BK \cite{baiwei2023}), KM (e.g., BO-KM \cite{bai2025}, KUPDAP \cite{Sugiyama2015}), and arbitrary distributions (e.g., LEOPARD \cite{Astfalk2017}, ALPS \cite{Verscharen2018_ALPS, Klein2025}, BO-Arbitrary \cite{xie2025}).
Listed in Table \ref{tab:BO-familyList}, the BO-family solvers employ a matrix method to model plasmas with various distributions (BM, KM, PBK, and arbitrary gyrotropic, incorporating parallel, perpendicular drift, ring-beam, and loss-cone features). The BO-Arbitrary code \cite{xie2025} expands the distribution function using Hermite basis functions and transforms the dispersion relation into a matrix eigenvalue problem via rational approximation, though its fitting accuracy for low-order kappa distributions is limited.

Building on these advances, this paper presents the BO-PBK solver, a new eigenvalue-based approach for analyzing waves and instabilities of obliquely propagating waves in magnetized multi-component plasmas. The solver supports drifting distributions (PBK, KM, BM), and their loss-cone hybrids. The method reformulates the generalized dielectric tensor for such plasmas as a linear eigenvalue system. This reformulation eliminates the need for initial-value iteration and enables the accurate, one-time calculation of all significant eigenroots. Furthermore, this paper elaborates on the eigenvalue system construction of the solver and presents numerical verification. 
\begin{table}[ht]
\caption{List of BO-family solvers}
\centering
\setlength{\tabcolsep}{3pt}
\begin{tabular}{cccccc}
\hline\noalign{\smallskip}\hline\noalign{\smallskip}
\textbf{Solver} & \textbf{PDRK} \cite{xie2016} 
& \textbf{BO} \cite{xie2019} & \textbf{BO-Arbitray} \cite{xie2025}
& \textbf{BO-KM} \cite{bai2025} & \textbf{BO-PBK} \\
\noalign{\smallskip}\hline\noalign{\smallskip} 
\textbf{Distribution}
& \makecell{BM \\ (Drift)}  
& \makecell{BM \\ (Loss-cone, \\ Drift, \\Ring beam)}
& Arbitrary 
& \makecell{KM/BM \\ (Drift)}
& \makecell{PBK/KM/BM \\ (Loss-cone, \\ Drift)}
\\
\noalign{\smallskip} \hline\hline\noalign{\smallskip}
\end{tabular}
\label{tab:BO-familyList}       
\parbox{\linewidth}{\raggedright \scriptsize
\textit{KM/BM: KM, BM, and their multi-component mixed-distribution plasmas; PBK/KM/BM: PBK, KM, BM, and their multi-component mixed-distribution plasmas.}
}
\end{table}

This paper is organized as follows. Section \ref{sec2} introduces the models of PBK-Maxwellian plasma distribution models and derives the concise dispersion tensors. Section \ref{sec3} details the construction of the eigenvalue system for the solver. Section \ref{sec4} presents benchmark studies for the new solver. Finally, Section \ref{sec5} concludes and discusses the findings.


\section{Hybrid kinetic model for PBK-Maxwellian plasmas}
\label{sec2}
Based on the Vlasov-Maxwell equations, we derive a concise rational expression for the susceptibility tensor of a PBK plasma. This formulation reduces to the well-established rational forms of KM and BM susceptibility tensors in the Maxwellian limit, thereby providing a unified description across these distribution models.

\subsection{PBK plasma distribution functions}
This study considers plasma particles governed by loss-cone drift PBK distributions with direction-dependent indices $\kappa_{\parallel s}$ and $\kappa_{\perp s}$ for the $s$th species, given by
\begin{eqnarray}
\label{PBKs}
\begin{cases}
f_{s0}(v_{\parallel}, v_{\perp}) 
= f_{s0}^\mathrm{PBK}(v_{\parallel}, v_{\perp}^2) 
= f_{s\parallel}^\mathrm{K}(v_{\parallel}) f_{s\perp}^\mathrm{K}(v_{\perp}^2), 
\\
f_{s\parallel}^\mathrm{K}(v_{\parallel}) =  \frac{1}{\sqrt{\pi} \theta_{\parallel s}}
\frac{\Gamma(\kappa_{\parallel s}+1)}{\sqrt{\kappa_{\parallel s}}\Gamma(\kappa_{\parallel s}+1/2)}
\left[1 + \frac{(v_{\parallel} - u_{s0})^2}{\kappa_{\parallel s}\theta_{\parallel s}^2}\right]^{-\left(\kappa_{\parallel s} +1\right)},
\\
f_{s\perp}^\mathrm{K}(v_{\perp}^2) = 
\frac{1}{\pi \theta_{\perp s}^2 \kappa_{\perp s}^{\sigma_s +1}} 
\frac{\Gamma(\kappa_{\perp s} + \sigma_s + 1)}{\Gamma(\sigma_{s}+1)\Gamma(\kappa_{\perp s})}
\left(\frac{v_\perp}{\theta_{\perp s}} \right)^{2\sigma_s}
\left(1 + \frac{v_{\perp}^2}{\kappa_{\perp s}\theta_{\perp s}^2}\right)^{-\left(\kappa_{\perp s}+\sigma_{s}+1\right)},
\end{cases}
\end{eqnarray}
in polar coordinates $(v_x, v_y, v_z) = (v_\perp \cos\phi, v_\perp \sin\phi, v_\parallel)$ of the particle velocity space, where $v_\parallel$ and $v_\perp$ are the velocity components parallel and perpendicular to the background magnetic field  $\boldsymbol{B}_0$, respectively. 
The PBK distribution is normalized to unity, i.e., $\int f_{s0} d^3v = 1$. The parallel and perpendicular thermal velocities, $\theta_{\parallel s}$ and $\theta_{\perp s}$, are defined by 
\begin{equation}
\theta_{\parallel s} = \left[\frac{(2-1/\kappa_{\parallel s})k_B T_{\parallel s}}{m_s}\right]^{1/2},
\quad
\theta_{\perp s} = \left[
\frac{2(1-1/\kappa_{\perp s})k_B T_{\perp s}}{m_s \left(\sigma_s +1\right)}\right]^{1/2},
\end{equation}
with the indices constrained by $\kappa_{\parallel s} > 1/2$ and $\kappa_{\perp s} >1$, respectively. Here $k_B$ is Boltzmann's constant and $m_s$ denotes particle mass, and the field-aligned drift velocity is $u_{s0} = \int_{-\infty}^\infty f_{s\parallel}^\mathrm{K} v_{\parallel} d v_{\parallel}$. The PBK distribution used here (Eqs. (\ref{PBKs})) has an exponent differing by one from that in Ref. \cite{bai2024}, which results in a corresponding unit offset in the kappa index of the dispersion relation, as confirmed by both analytical and numerical results.

\begin{figure}[t]
\centering
\includegraphics[scale=0.34]{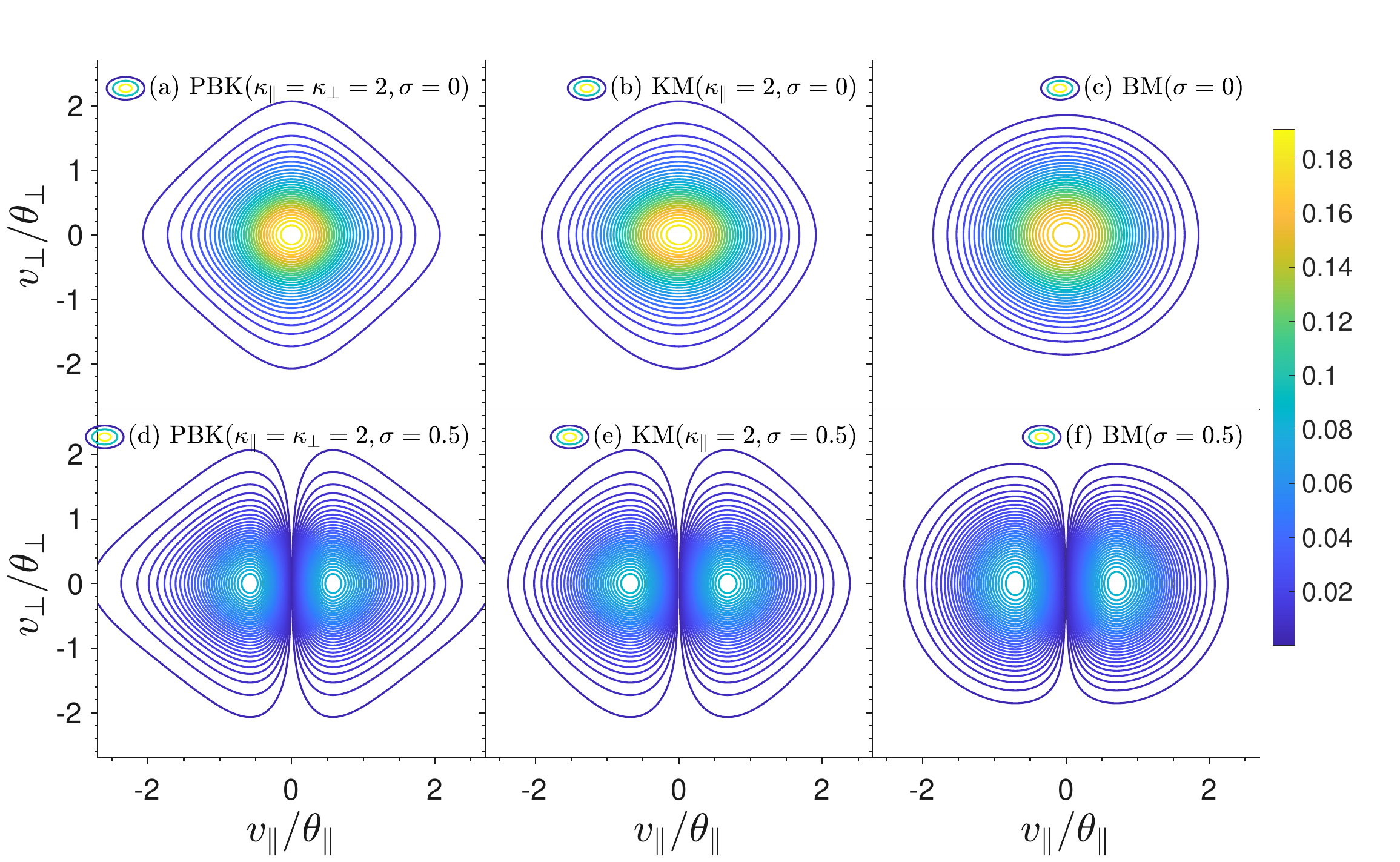}
\caption{Contours of the normalized electron velocity distribution functions in the ($v_{\parallel}$, $v_{\perp}$) plane with $T_{\parallel e} = T_{\perp e} = 1\times 10^2$ eV. Panels (a)-(c) show the PBK, KM, and BM distributions without a loss-cone, while panels (d)-(f) show the corresponding distributions with a loss-cone $\sigma = 0.5$.}
\label{VDFs_contour}
\end{figure}

\begin{figure}[t]
\centering
\includegraphics[scale=0.4]{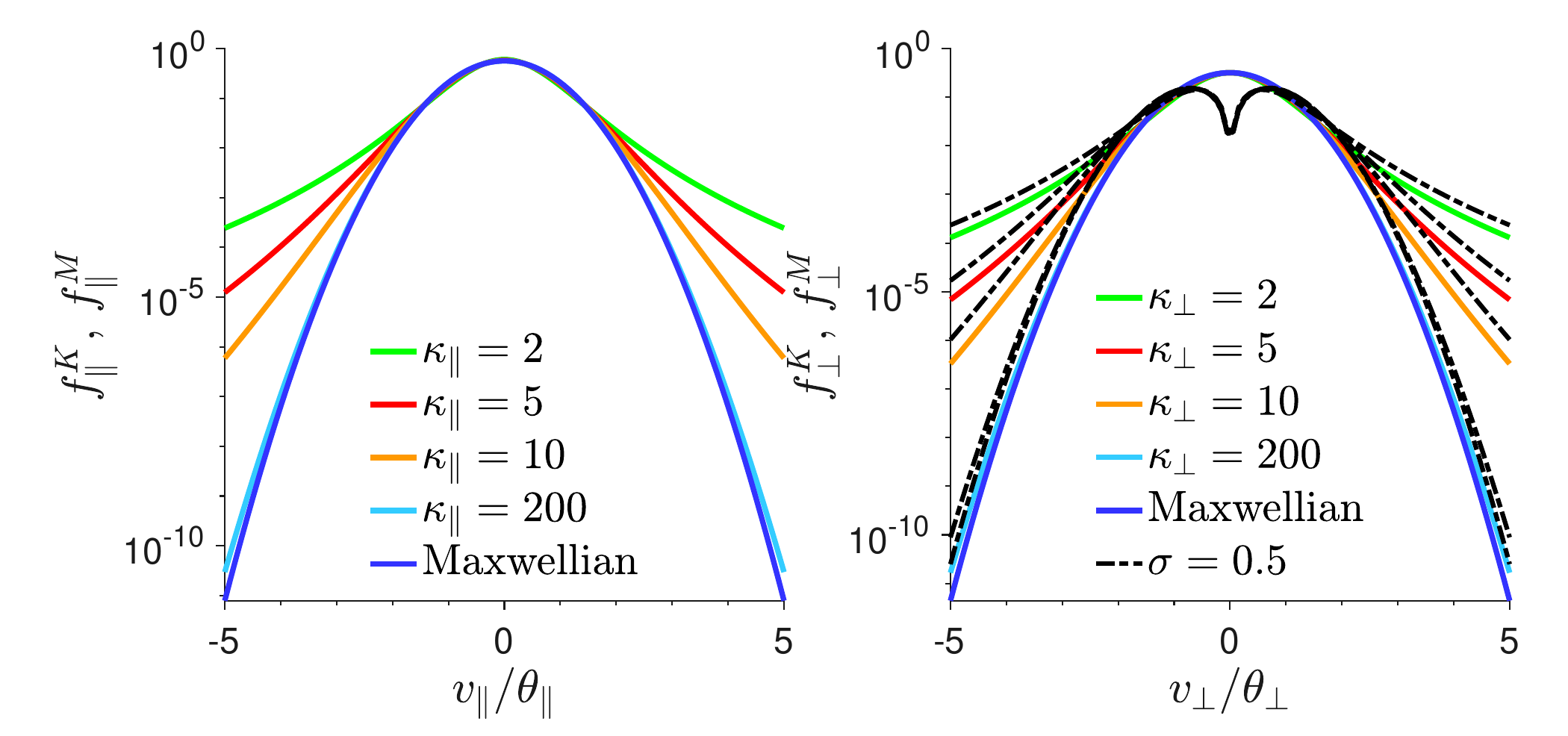}
\caption{Normalized electron velocity distribution functions for Kappa and Maxwellian models at $T_{\parallel e} = T_{\perp e} = 1\times 10^2$ eV. Left: parallel distributions $f_{\parallel}^\mathrm{K}(v_{\parallel})$ and $f_{\parallel}^\mathrm{M}(v_{\parallel})$; right: perpendicular distributions $f_{\perp}^\mathrm{K}(v_{\perp})$ and $f_{\perp}^\mathrm{M}(v_{\perp})$ with a loss-cone at $\sigma=0.5$.}
\label{VDF1D_pbk}
\end{figure}

In Maxwellian limit ($\kappa_{\perp s}\rightarrow \infty$), the loss-cone drift PBK distribution (\ref{PBKs}) transforms into the loss-cone drift KM distribution function, given by
\begin{eqnarray}
\label{KMs}
f_{s0}^\mathrm{KM}(v_{\parallel}, v_{\perp}^2) 
= \lim_{\kappa_{\perp s}\rightarrow \infty} f_{s0}^\mathrm{PBK}(v_{\parallel}, v_{\perp}^2) 
= f_{s\parallel}^\mathrm{K}(v_{\parallel}) f_{s\perp}^\mathrm{M}(v_{\perp}^2).
\end{eqnarray}
Furthermore, when the loss-cone drift PBK distribution (\ref{PBKs}) is in Maxwellian limits ($\kappa_{\parallel s}\rightarrow \infty$ and $\kappa_{\perp s}\rightarrow \infty$), it takes the form of the loss-cone drift BM distribution, denoted as
\begin{eqnarray}
\label{BMs}
f_{s0}^\mathrm{BM}(v_{\parallel}, v_{\perp}^2) 
= \lim_{\kappa_{\parallel s},\kappa_{\perp s}\rightarrow \infty}
f_{s0}^\mathrm{PBK}(v_{\parallel}, v_{\perp}^2) 
= f_{s\parallel}^\mathrm{M}(v_{\parallel}) f_{s\perp}^\mathrm{M}(v_{\perp}^2).
\end{eqnarray}
Here, $f_{s,\parallel}^\mathrm{M}$ and $f_{s,\perp}^\mathrm{M}$ represent the Maxwellian distribution functions for velocity components parallel and perpendicular to the magnetic field, respectively. They can be expressed as follows,
\begin{eqnarray}
\begin{cases}
f_{s\parallel}^\mathrm{M}(v_{\parallel})  
&= \lim_{\kappa_{\parallel s}\rightarrow \infty}
f_{s\parallel}^\mathrm{K}(v_{\parallel}) 
= \frac{1}{\sqrt{\pi} \tilde{\theta}_{\parallel s}}
\exp\left[-\frac{\left(v_{\parallel} - u_{s0}\right)^2}{\tilde{\theta}_{\parallel s}^2}\right],
\\
f_{s\perp}^\mathrm{M}(v_{\perp}^2)  
&= \lim_{\kappa_{\perp s}\rightarrow \infty}
f_{s\perp}^\mathrm{K}(v_{\perp}^2) 
= \frac{1}{\pi \tilde{\theta}_{\perp s}^2 \Gamma(\sigma_s+1)} 
\left(\frac{v_\perp}{\tilde\theta_{\perp s}} \right)^{2\sigma_s}
\exp\left(- \frac{v_{\perp}^2}{\tilde{\theta}_{\perp s}^2}\right),
\end{cases}
\end{eqnarray}
where $\tilde{\theta}_{\parallel s}$ and $\tilde{\theta}_{\perp s}$ denote the thermal speeds parallel and perpendicular to the magnetic field in Maxwellian plasma, respectively, and are defined as:
\begin{equation}
\tilde{\theta}_{\parallel s} = \left(\frac{2 k_B T_{\parallel s}}{m_s}\right)^{1/2},
\quad
\tilde{\theta}_{\perp s} =  \left[\frac{2 k_B T_{\perp s}}{m_s\left(\sigma_s + 1\right)}\right]^{1/2}.
\end{equation}
Figure \ref{VDFs_contour} shows the contour of normalized electron velocity distributions in the ($v_{\parallel}$, $v_{\perp}$) planes. Panels (a)-(c) correspond to the PBK, KM, and BM models without a loss-cone,  while panels (d)-(f) show the same models with a loss-cone ($\sigma = 0.5$). Figure \ref{VDF1D_pbk} plots the normalized parallel distributions $f_{\parallel}^\mathrm{K}(v_{\parallel})$ and $f_{\parallel}^\mathrm{M}(v_{\parallel})$ (left), and the perpendicular distributions $f_{\perp}^\mathrm{K}(v_{\perp})$ and $f_{\perp}^\mathrm{M}(v_{\perp})$ with $\sigma=0.5$ (right), revealing that the distribution functions converge toward the Maxwellian distribution with increasing $\kappa_{\parallel}$ and $\kappa_{\perp}$, and are close to it at $\kappa_{\parallel}=\kappa_{\perp}=200$.

\subsection{The dispersion tensor for PBK plasma}
We consider an obliquely propagating plasma wave ($\propto\exp[i(\boldsymbol{k}\cdot\boldsymbol{r} - \omega t)]$), where the wavevector $\boldsymbol{k}=(k_x, 0,k_z) = (k\sin\theta, 0, k\cos\theta)$ forms an angle $\theta$ with the background magnetic field $\boldsymbol{B}_0=B_0\boldsymbol{e}_z$. The susceptibility tensor $\boldsymbol{\chi}_s$ for a magnetized plasma, derived from the Vlasov-Maxwell equations, contributes to the dielectric tensor as $\boldsymbol{\epsilon (\omega, \boldsymbol{k})} = \boldsymbol{I} + \sum_s \boldsymbol{\chi}_s$,  where $\boldsymbol{I}$ is the unit dyadic and the sum is over all plasma species. The full susceptibility $\boldsymbol{\chi} = \sum_s\boldsymbol{\chi}_s$ is \cite{stix1992waves,Summers1994,summers_Thorne1995}, 
\begin{eqnarray}
\begin{split}
\boldsymbol{\chi}
= \sum_s \sum_{n=-\infty}^{\infty}
\frac{\omega^2_{ps}}{\omega} \int 
\frac{v_{\perp} d\boldsymbol{v}}
{\omega - k_{\parallel} v_{\parallel} - n\omega_{cs}}
\times
\left[
\begin{array}{ccc}
A_s \frac{n^2}{\mu_s^2} J_{n}^2  
& i A_s\frac{n}{\mu_s}  J_n  J_n^{\prime}   
& B_s \frac{n}{\mu_s} J_{n}^2  
\\
- i A_s\frac{n}{\mu_s}  J_n  J_n^{\prime}   
& A_s J_n^{\prime 2}  
& -i B_s J_n  J_n^{\prime}  
\\
A_s \frac{n}{\mu_s} \frac{v_{\parallel}}{v_{\perp}} J_n^2 
& i A_s \frac{v_{\parallel}}{v_{\perp}} J_n J_n^{\prime}
& B_s \frac{v_{\parallel}}{v_{\perp}} J_n^{2} 
\end{array}
\right],
\end{split}
\end{eqnarray}
where the velocity-space integral is $\int \left(\cdots \right) d\boldsymbol{v} = \int_{-\infty}^{\infty} \left(\cdots \right) d v_{\parallel} \int_{0}^{\infty} 2 \pi v_{\perp} d v_{\perp}$, $J_{n}=J_{n}(\mu_s)$ is the Bessel function of the first kind of order $n$ (for $n=0,\pm1, \pm2,\cdots$), and $J_{n}^{\prime}=dJ_n(\mu_s)/d\mu_s$ denotes its derivative. The argument $\mu_s$ is given by $\mu_s = k_{\perp} v_{\perp}/\omega_{cs}$, with the gyrofrequency $\omega_{cs} = q_s B_0/m_{s}$, and plasma frequency $\omega_{ps} =\sqrt{n_{0s} q_{s}^2/(\epsilon_0 m_{s}})$. Here, $q_s$, $m_{s}$ and $n_{0s}$ are particle charge, mass and number density, respectively. We define the $A_s $ and $B_s$ by
\begin{subequations}
\label{As_Bs_KDR}
\begin{align}
\label{As_KDR}
&A_s \equiv
\left[\left(1 - \frac{k_{\parallel} v_{\parallel} }{\omega}\right)
\frac{\partial }{\partial v_{\perp}} 
+ \frac{k_{\parallel} v_{\perp} }{\omega} \frac{\partial }{\partial v_{\parallel}}
\right] f_{s0},
\\
\label{B_s_KDR}
&B_s \equiv
\left[\frac{n \omega_{cs}}{\omega } \left(\frac{v_{\parallel}}{v_{\perp}}\right)
\frac{\partial }{\partial v_{\perp}} 
+ \left(1-  \frac{n \omega_{cs} }{\omega} \right) \frac{\partial }{\partial v_{\parallel}}
\right] f_{s0}.
\end{align}
\end{subequations}
To further derive the expression for the susceptibility $\boldsymbol{\chi}$, we use the modified plasma dispersion function introduced by Summers and Thorne \cite{Summers1991}, defined as
\begin{equation}
Z_{\kappa_{\parallel s}}(\xi_n) =
\frac{\Gamma(\kappa_{\parallel s}+1)}{\sqrt{\pi\kappa_{\parallel s}^3}\Gamma(\kappa_{\parallel s}-1/2)}
\int_{-\infty}^{\infty} 
\frac{dx}{\left(x-\xi_n\right)\left(1+ x^2/\kappa_{\parallel s} \right)^{\kappa_{\parallel s}+1}},
\quad \Im(\xi)>0,
\end{equation}
with $x = (v_{\parallel} - u_{s0})/\theta_{\parallel s}$ and
$\xi_n = (\omega - n\omega_{cs} - k_{\parallel} u_{s0})/k_{\parallel} \theta_{\parallel s}$.
For any positive integer $\kappa_{\parallel s}$, and all $\xi_n \neq \pm i \sqrt{\kappa_{\parallel s}}$, the expressions for these dispersion functions can be given using closed-form expansions \cite{baiwei2023, Summers1994},
\begin{equation}
\label{Z_kappa_closed-form}
Z_{\kappa_{\parallel s}}(\xi_n) 
= \frac{i(\kappa_{\parallel s}-1/2)}{2\kappa_{\parallel s}^{3/2}}
\frac{\kappa_{\parallel s}!}{(2\kappa_{\parallel s})!} 
\sum_{l=1}^{\kappa_{\parallel s}+1} \frac{(\kappa_{\parallel s}+l-1)!}{(l-1)!} 
\left(\frac{2i }{(\xi_n /\sqrt{\kappa_{\parallel s}})+i} \right)^{\kappa_{\parallel s}-l+2}.
\end{equation}
We define $\mathcal Z_{\kappa_{\parallel s}} \equiv
Z_{\kappa_{\parallel s}}/\left[1-1/(2\kappa_{\parallel s})\right]$ and 
$\mathcal Z_{\kappa_{\parallel s}}^{\prime}\equiv \partial \mathcal Z_{\kappa_{\parallel s}}/\partial \xi_n$. The closed-form expressions are given by
\begin{subequations}
\label{Z_kappa_form2}
\begin{align}
\label{def_Z_kappa_closed-form2}
\begin{split}
&\mathcal Z_{\kappa_{\parallel s}} 
= \frac{i}{2\sqrt{\kappa_{\parallel s}}}
\frac{\kappa_{\parallel s}!}{(2\kappa_{\parallel s})!} 
\sum_{l=1}^{\kappa_{\parallel s}+1} \frac{(\kappa_{\parallel s}+l-1)!}{(l-1)!} 
\left(\frac{2i }{(\xi_n /\sqrt{\kappa_{\parallel s}})+i} \right)^{\kappa_{\parallel s}-l+2},
\end{split}
\\
\label{def_dZ_kappa_closed-form2}
\begin{split}
& \mathcal Z_{\kappa_{\parallel s}}^{\prime}
= -\frac{\left(\kappa_{\parallel s}+1/2\right)}{\kappa_{\parallel s}}
\frac{\left(\kappa_{\parallel s}+1\right)!}{(2\kappa_{\parallel s}+2)!} 
\sum_{l=1}^{\kappa_{\parallel s}+1} 
\left(\kappa_{\parallel s}-l+2\right)
\frac{(\kappa_{\parallel s}+l-1)!}{(l-1)!} 
\left( \frac{2 i}{(\xi_n /\sqrt{\kappa_{\parallel s}})+i} \right)^{\kappa_{\parallel s}-l+3}.
\end{split}
\end{align}
\end{subequations}
Following the derivation, the susceptibility componetns $\chi_{ij}$ for a PBK plasma are
\begin{subequations}
\label{chi_PBK}
\begin{align}
\label{ch11_PBK}
&\chi_{11}^\mathrm{PBK} = -\sum_{s} \frac{\omega^2_{ps}}{\omega^2}
+ \sum_{sn} \frac{\omega^2_{ps}}{\omega^2} n^2 C_{12sn}^\mathrm{PBK},
\\
\label{ch12_PBK}
&\chi_{12}^\mathrm{PBK} =  i\sum_{sn} \frac{\omega^2_{ps}}{\omega^2} n C_{34sn}^\mathrm{PBK},
\\
\label{ch13_PBK}
&\chi_{13}^\mathrm{PBK} = \tan \theta \left[\sum_{s} \frac{\omega^2_{ps}}{\omega^2} 
+ \sum_{sn} \frac{\omega^2_{ps}}{\omega^2} n
\left(\frac{\omega-n\omega_{cs}}{\omega_{cs}}\right) C_{12sn}^\mathrm{PBK}\right],
\\
\label{ch22_PBK}
&\chi_{22}^\mathrm{PBK} = -\sum_{s} \frac{\omega^2_{ps}}{\omega^2} 
+ \sum_{sn} \frac{\omega^2_{ps}}{\omega^2} C_{56sn}^\mathrm{PBK},
\\
\label{ch23_PBK}
&\chi_{23}^\mathrm{PBK} = -i \tan \theta \sum_{sn} \frac{\omega^2_{ps}}{\omega^2} 
\left(\frac{\omega-n\omega_{cs}}{\omega_{cs}}\right) C_{34sn}^\mathrm{PBK},
\\
\label{ch33_PBK}
&\chi_{33}^\mathrm{PBK} = \tan^2 \theta \left[-\sum_{s} \frac{\omega^2_{ps}}{\omega^2} 
+ \sum_{sn} \frac{\omega^2_{ps}}{\omega^2} 
\left(\frac{\omega - n\omega_{cs}}{\omega_{cs}}\right)^2 C_{12sn}^\mathrm{PBK}\right],
\end{align}
\end{subequations}
where $\chi_{12}^\mathrm{PBK} = -\chi_{21}^\mathrm{PBK}$, $\chi_{13}^\mathrm{PBK} = \chi_{31}^\mathrm{PBK}$, and $\chi_{23}^\mathrm{PBK} = -\chi_{32}^\mathrm{PBK}$. 
Using the closed forms in Eqs. (\ref{def_Z_kappa_closed-form2}) and (\ref{def_dZ_kappa_closed-form2}), the coefficients $C_{12sn}^\mathrm{PBK}$, $C_{34sn}^\mathrm{PBK}$, and $C_{56sn}^\mathrm{PBK}$ take the simplified rational form
\begin{subequations}
\label{Cn_pbk_rational_expansion}
\begin{align}
\label{Cn12sn_rational_expansion}
C_{12sn}^\mathrm{PBK} 
=\left(\frac{n \omega_{cs}}{k_{\parallel} \theta_{\parallel s}}\right) S_{17} 
\mathcal Z_{\kappa_{\parallel s}}
-\left(\frac{\theta_{\perp s}^2 }{2\theta_{\parallel s}^2}\right)  S_2  
\mathcal Z_{\kappa_{\parallel s}}^{\prime}
= \sum_{l=1}^{\kappa_{\parallel s}} 
\left[\frac{b_{1snl}}{\left(\omega - c_{sn}\right)^l}
+ \frac{b_{2snl}}{\left(\omega - c_{sn}\right)^{l+1}} \right],
\\
\label{Cn34sn_rational_expansion}
C_{34sn}^\mathrm{PBK} 
= \left(\frac{n \omega_{cs}}{k_{\parallel} \theta_{\parallel s}} \right) S_{38}  
\mathcal Z_{\kappa_{\parallel s}}
-\left(\frac{\theta_{\perp s}^2}{2\theta_{\parallel s}^2} \right) S_4  
\mathcal Z_{\kappa_{\parallel s}}^{\prime}
= \sum_{l=1}^{\kappa_{\parallel s}} 
\left[ \frac{b_{3snl}}{\left(\omega - c_{sn}\right)^l}
+ \frac{b_{4snl}}{\left(\omega - c_{sn}\right)^{l+1}} \right],
\\
\label{Cn56sn_rational_expansion}
C_{56sn}^\mathrm{PBK} 
= \left(\frac{n\omega_{cs}}{k_{\parallel} \theta_{\parallel s}}\right) S_{59} 
\mathcal Z_{\kappa_{\parallel s}}
-\left(\frac{\theta_{\perp s}^2}{2\theta_{\parallel s}^2}\right) S_6  
\mathcal Z_{\kappa_{\parallel s}}^{\prime}
= \sum_{l=1}^{\kappa_{\parallel s}} 
\left[ \frac{b_{5snl}}{\left(\omega - c_{sn}\right)^l}
+ \frac{b_{6snl}}{\left(\omega - c_{sn}\right)^{l+1}} \right].
\end{align}
\end{subequations}
where,
\begin{eqnarray}
\begin{cases}
\begin{split}
&b_{1snl} = S_{17} b_{snl}, 
\quad
b_{3snl} = S_{38} b_{snl}, 
\quad
b_{5snl} = S_{59} b_{snl},
\\
&b_{2snl} = S_2 b_{sl},
\quad
b_{4snl} = S_4 b_{sl},
\quad
b_{6snl} = S_6 b_{sl},
\\
&S_{17} = \left(\frac{\kappa_{\perp s}+\sigma_s+1}{\kappa_{\perp s}}\right)S_1 
-2\sigma_s \lambda_s S_7,
\\
&S_{38} = \left(\frac{\kappa_{\perp s}+\sigma_s+1}{\kappa_{\perp s}}\right)S_3 
- 2\sigma_s \lambda_s S_8,
\\
&S_{59} = \left(\frac{\kappa_{\perp s}+\sigma_s+1}{\kappa_{\perp s}}\right)S_5
- 2\sigma_s \lambda_s S_9,
\\
&b_{sl} = -\frac{l}{2} k_{\parallel}^2 \theta_{\perp s}^2 c_{sl},
\quad
b_{snl} = -n\omega_{cs} c_{sl},
\\
&c_{sn} = n\omega_{cs} + k_{\parallel}u_{s0} - i \sqrt{\kappa_{\parallel s}} k_\parallel \theta_{\parallel s},
\\
&c_{sl} 
= \frac{\Gamma(\kappa_{\parallel s}+1)\Gamma(2\kappa_{\parallel s}-l+2)}{
\Gamma(2\kappa_{\parallel s}+1) \Gamma(\kappa_{\parallel s}-l+2)}
\left(2i\sqrt{\kappa_{\parallel s}} k_\parallel \theta_{\parallel s}\right)^{l-1},
\end{split}
\end{cases}
\end{eqnarray}
and $\left(c_{sl}\right)_{l=1}=1$. The unified integral expression is given by
\begin{subequations}
\begin{align}
\label{S_kappa_perp}
&S_{\kappa_{\perp s}}\left(p_1,p_2,p_3,p_4\right) 
= S_0 \int_{0}^{\infty} 
\frac{J_{n}^{p_1}(\mu_s) J_{n}^{\prime p_2}(\mu_s) \mu_s^{2\sigma_s+p_3}}{\left(1 + \frac{\mu_s^2}{2\lambda_s \kappa_{\perp s}}\right)^{\kappa_{\perp s}+\sigma_s+p_4}} 
d\mu_s,
\\
&S_0 = \frac{4\Gamma(\kappa_{\perp s} + \sigma_s + 1)}
{\left(2\lambda_s\right)^{\sigma_s+2}\left(\kappa_{\perp s}\right)^{\sigma_s+1}
\Gamma(\kappa_{\perp s})\Gamma(\sigma_s + 1)},
\end{align}
\end{subequations}
where $\lambda_{s} = \frac{k_{\perp}^2 \theta_{\perp s}^2}{2\omega_{cs}^2}$, $J_{n}^{p_1}(\mu_s) = \left[J_{n}(\mu_s)\right]^{p_1}$, and $J_{n}^{\prime p_2}(\mu_s) = \left[\frac{dJ_{n}(\mu_s)}{d\mu_s}\right]^{p_2}$. In terms of this expression, the nine integrals in equation set (\ref{Cn_pbk_rational_expansion}) can be compactly written as
\begin{eqnarray}
\begin{cases}
S_1= S_{\kappa_{\perp s}}\left(2,0,1,2\right),
\quad
S_2= S_{\kappa_{\perp s}}\left(2,0,1,1\right),
\quad
S_3= S_{\kappa_{\perp s}}\left(1,1,2,2\right),
\\
S_4= S_{\kappa_{\perp s}}\left(1,1,2,1\right),
\quad
S_5= S_{\kappa_{\perp s}}\left(0,2,3,2\right),
\quad
S_6= S_{\kappa_{\perp s}}\left(0,2,3,1\right),
\\
S_7= S_{\kappa_{\perp s}}\left(2,0,-1,1\right),
\quad
S_8= S_{\kappa_{\perp s}}\left(1,1,0,1\right),
\quad
S_9= S_{\kappa_{\perp s}}\left(0,2,1,1\right).
\end{cases}
\end{eqnarray}
When computing the integrals in Eq. (\ref{S_kappa_perp}) numerically, care must be taken to prevent numerical overflow due to large value of $\kappa_{\perp s}$, for enhanced stability, the coefficient $S_0$ should be evaluated in its logarithmic form
$S_0 = \frac{4\exp\left[\ln\Gamma(\kappa_{\perp s} + \sigma_s + 1) - \ln\Gamma(\kappa_{\perp s})\right]}
{\left(2\lambda_s\right)^{\sigma_s+2} \left(\kappa_{\perp s}\right)^{\sigma_s+1} 
\Gamma(\sigma_s + 1)}$. Handing the $\lambda_s \rightarrow 0$ limit requires a change of the integration variable $\hat{\mu} = \mu/\sqrt{2\lambda_s}$ \cite{Summers1994} for numerical stability, as the integral (\ref{S_kappa_perp}) becomes critical when $k_{\perp} \rightarrow 0$ or $\theta_{\perp s} \rightarrow 0$. This yields the transformed integral
\begin{equation}
\label{S_kappa_perp_transformation}
S_{\kappa_{\perp s}}\left( p_1,p_2,p_3,p_4\right) 
= \hat S_0\int_{0}^{\infty}
\frac{J_{n}^{p_1}(\sqrt{2\lambda_s}\hat\mu)
J_{n}^{\prime p_2}(\sqrt{2\lambda_s}\hat\mu)\hat\mu^{2\sigma_s+p_3}}
{\left(1 + \frac{\hat\mu^2}{ \kappa_{\perp s}}\right)^{\kappa_{\perp s}+\sigma_s+p_4}} d \hat\mu,
\end{equation}
where $\hat S_0 = \frac{4\exp\left[\ln\Gamma(\kappa_{\perp s} + \sigma_s + 1) 
-\ln\Gamma(\kappa_{\perp s})\right]}
{\left(2\lambda_s\right)^{\left(3-p_3\right)/2}
\left(\kappa_{\perp s}\right)^{\sigma_s+1}\Gamma(\sigma_s + 1)}$.

\subsection{Susceptibility tensors in the Maxwellian limit}
In the Maxwellian limit ($\kappa_{\perp s}\rightarrow \infty$), the coefficients given by equation set (\ref{Cn_pbk_rational_expansion}) become
\begin{eqnarray}
\label{Cn_KM}
\begin{cases}
\begin{split}
\lim_{\kappa_{\perp s}\rightarrow\infty} C_{12sn}^\mathrm{PBK}
= \left(\frac{n \omega_{cs}}{k_{\parallel} \theta_{\parallel s}}\right) 
\tilde S_{17} \mathcal Z_{\kappa_{\parallel s}}
-\left(\frac{\theta_{\perp s}^2 }{2\theta_{\parallel s}^2}\right) 
\tilde S_2 \mathcal Z_{\kappa_{\parallel s}}^{\prime}
=
\sum_{l=1}^{\kappa_{\parallel s}} 
\left[\frac{\tilde b_{1snl}}{\left(\omega - c_{sn}\right)^l}
+ \frac{\tilde b_{2snl}}{\left(\omega - c_{sn}\right)^{l+1}} \right],
\\
 \lim_{\kappa_{\perp s}\rightarrow\infty} C_{34sn}^\mathrm{PBK}
= \left(\frac{n \omega_{cs}}{k_{\parallel} \theta_{\parallel s}} \right)
\tilde S_{38} \mathcal Z_{\kappa_{\parallel s}}
-\left(\frac{\theta_{\perp s}^2}{2\theta_{\parallel s}^2} \right)
\tilde S_4 \mathcal Z_{\kappa_{\parallel s}}^{\prime}
= \sum_{l=1}^{\kappa_{\parallel s}} 
\left[ \frac{\tilde b_{3snl}}{\left(\omega - c_{sn}\right)^l}
+ \frac{\tilde b_{4snl}}{\left(\omega - c_{sn}\right)^{l+1}} \right],
\\
\lim_{\kappa_{\perp s}\rightarrow\infty} C_{56sn}^\mathrm{PBK}
= \left(\frac{n\omega_{cs}}{k_{\parallel} \theta_{\parallel s}}\right)
\tilde S_{59} \mathcal Z_{\kappa_{\parallel s}}
-\left(\frac{\theta_{\perp s}^2}{2\theta_{\parallel s}^2}\right) 
\tilde S_6 \mathcal Z_{\kappa_{\parallel s}}^{\prime}
= \sum_{l=1}^{\kappa_{\parallel s}} 
\left[ \frac{\tilde b_{5snl}}{\left(\omega - c_{sn}\right)^l}
+ \frac{\tilde b_{6snl}}{\left(\omega - c_{sn}\right)^{l+1}} \right],
\end{split}
\end{cases}
\end{eqnarray}
where
\begin{eqnarray}
\label{coef_KM}
\begin{cases}
\begin{split}
&\tilde b_{1snl} = \tilde S_{17} b_{snl}, 
\quad
\tilde b_{3snl} = \tilde S_{38} b_{snl}, 
\quad
\tilde b_{5snl} = \tilde S_{59} b_{snl}, 
\\
&\tilde b_{2snl} = \tilde S_2 b_{sl},
\quad
\tilde b_{4snl} = \tilde S_4 b_{sl},
\quad
\tilde b_{6snl} = \tilde S_6 b_{sl},
\quad
\tilde S_0 = \frac{4}{\left(2\lambda_s\right)^{\sigma_s+2} \Gamma(\sigma_s+1)},
\\
&\tilde S_1 = \tilde S_2
= \tilde S_0 \int_{0}^{\infty} J_{n}^2\mu_s^{2\sigma_s + 1}
\exp\left(-\frac{\mu_s^2}{2\lambda_s}\right) d\mu_s,
\quad
\tilde S_1\big|_{\sigma_s=0} = \tilde S_2\big|_{\sigma_s=0} = \Lambda_n/\lambda_s,
\\
&\tilde S_3 = \tilde S_4
= \tilde S_0 \int_{0}^{\infty} J_{n} J_{n}^{\prime}\mu_s^{2\sigma_s + 2}
\exp\left(-\frac{\mu_s^2}{2\lambda_s}\right) d\mu_s,
\quad
\tilde S_3\big|_{\sigma_s=0} = \tilde S_4\big|_{\sigma_s=0} = \Lambda_n^{\prime},
\\
&\tilde S_5 = \tilde S_6 
= \tilde S_0 \int_{0}^{\infty} J_{n}^{\prime 2}\mu_s^{2\sigma_s + 3}
\exp\left(-\frac{\mu_s^2}{2\lambda_s}\right) d\mu_s,
\quad
\tilde S_5\big|_{\sigma_s=0} = \tilde S_6\big|_{\sigma_s=0}
= n^2 \Lambda_n/\lambda_s - 2\lambda_s\Lambda_n^{\prime},
\\
&\tilde S_7 = \tilde S_0 \int_{0}^{\infty} J_{n}^{2}\mu_s^{2\sigma_s - 1}
\exp\left(-\frac{\mu_s^2}{2\lambda_s}\right) d\mu_s,
\quad
\tilde S_8 = \tilde S_0 \int_{0}^{\infty} J_{n} J_{n}^{\prime}\mu_s^{2\sigma_s}
\exp\left(-\frac{\mu_s^2}{2\lambda_s}\right) d\mu_s,
\\
&\tilde S_9 = \tilde S_0 \int_{0}^{\infty} J_{n}^{\prime 2}\mu_s^{2\sigma_s + 1}
\exp\left(-\frac{\mu_s^2}{2\lambda_s}\right) d\mu_s,
\quad
\tilde S_{17} = \left(\tilde S_1 - 2\sigma_s \lambda_s \tilde S_7\right),
\\
&\tilde S_{38} = \left(\tilde S_3 - 2\sigma_s \lambda_s \tilde S_8\right),
\quad
\tilde S_{59} = \left(\tilde S_5 - 2\sigma_s \lambda_s \tilde S_9\right),
\end{split}
\end{cases}
\end{eqnarray}
with $\Lambda_n = I_n(\lambda_s)\exp(-\lambda_s)$ and $\Lambda_s^{\prime} = d\Lambda/d\lambda_s$,  where $I_n$ is the modified Bessel function of the first of order $n$. Substituting the coefficients from equation set (\ref{Cn_KM}) into the PBK susceptibility $\chi_{ij}^\mathrm{PBK}$ in equation set (\ref{chi_PBK}) yields the susceptibility components for the loss-cone drift KM plasma. In particular, for $\sigma_s = 0$, the relations for $\tilde S_1\big|_{\sigma_s=0}$ to $\tilde S_6\big|_{\sigma_s=0}$ from Eqs. (\ref{coef_KM}) simplify the result to 
\begin{eqnarray}
\label{Cn_max_sigma=0}
\begin{cases}
\begin{split}
\lim_{\kappa_{\perp s}\rightarrow\infty} C_{12sn}^\mathrm{PBK}\big|_{\sigma_s=0}
&= \frac{I_n e^{-\lambda_s}}{\lambda_s} C_n,
\\
\lim_{\kappa_{\perp s}\rightarrow\infty} C_{34sn}^\mathrm{PBK}\big|_{\sigma_s=0}
&= -\left(I_n - I_n^{\prime} \right)e^{-\lambda_s} C_n,
\\
\lim_{\kappa_{\perp s}\rightarrow\infty} C_{56sn}^\mathrm{PBK}\big|_{\sigma_s=0}
&= \frac{\left[n^2 I_n + 2\lambda_s^2 (I_n-I_n^{\prime})\right]e^{-\lambda_s}}{\lambda_s} C_n,
\end{split}
\end{cases}
\end{eqnarray}
where
$C_n = \left(\frac{n \omega_{cs}}{k_{\parallel} \theta_{\parallel s}}\right) 
\mathcal Z_{\kappa_{\parallel s}}
-\left(\frac{\theta_{\perp s}^2 }{2\theta_{\parallel s}^2}\right) 
\mathcal Z_{\kappa_{\parallel s}}^{\prime}$.
Substituting equation set (\ref{Cn_max_sigma=0}) into the general susceptibility $\chi_{ij}^\mathrm{PBK}$ in Eq. (\ref{chi_PBK}), yields a KM susceptibility tensor that reproduces the result in Refs. \cite{Cattaert2007, Sugiyama2015}:
\begin{eqnarray}
\label{chi_KM_sigma=0}
\begin{cases}
\begin{split}
\chi_{11}^\mathrm{KM}\big|_{\sigma_s=0} &= -\sum_{s} \frac{\omega^2_{ps}}{\omega^2}
+ \sum_{sn} \frac{\omega^2_{ps}}{\omega^2}
n^2 \frac{I_n  e^{-\lambda_s}}{\lambda_s} C_n,
\\
\chi_{12}^\mathrm{KM}\big|_{\sigma_s=0} &= -i\sum_{sn} \frac{\omega^2_{ps}}{\omega^2} 
n \left(I_n - I_n^{\prime} \right) e^{-\lambda_s} C_n,
\\
\chi_{13}^\mathrm{KM}\big|_{\sigma_s=0} &= \tan\theta\left[\sum_{s} \frac{\omega^2_{ps}}{\omega^2} 
+ \sum_{sn} \frac{\omega^2_{ps}}{\omega^2} 
\left(\frac{\omega-n\omega_{cs}}{\omega_{cs}}\right)
n \frac{I_n e^{-\lambda_s}}{\lambda_s} C_n \right],
\\
\chi_{22}^\mathrm{KM}\big|_{\sigma_s=0} &= -\sum_{s} \frac{\omega^2_{ps}}{\omega^2} 
+ \sum_{sn} \frac{\omega^2_{ps}}{\omega^2}
\frac{\left[n^2 I_n + 2\lambda_s^2 \left(I_n - I_n^{\prime} \right)\right] e^{-\lambda_s}}{\lambda_s} C_n,
\\
\chi_{23}^\mathrm{KM}\big|_{\sigma_s=0} &= i \tan \theta \sum_{sn} \frac{\omega^2_{ps}}{\omega^2}
\left(\frac{\omega-n\omega_{cs}}{\omega_{cs}}\right)
\left(I_n - I_n^{\prime} \right) e^{-\lambda_s} C_n,
\\
\chi_{33}^\mathrm{KM}\big|_{\sigma_s=0} &= \tan^2 \theta \left[-\sum_{s} \frac{\omega^2_{ps}}{\omega^2} 
+ \sum_{sn} \frac{\omega^2_{ps}}{\omega^2} 
\left(\frac{\omega - n\omega_{cs}}{\omega_{cs} }\right)^2
\frac{I_n  e^{-\lambda_s}}{\lambda_s} C_n \right].
\end{split}
\end{cases}
\end{eqnarray}

In Maxwellian limit ($\kappa_{\parallel s}\rightarrow \infty$), the modified plasma dispersion function converges to its standard form, $\lim_{\kappa_{\parallel s} \rightarrow \infty} \mathcal Z_{\kappa_{\parallel s}}(\xi) = Z(\xi)$, $\lim_{\kappa_{\parallel s} \rightarrow\infty} \mathcal Z^{\prime}_{\kappa_{\parallel s}}(\xi) = Z^{\prime}(\xi) = -2\left[1 + \xi Z(\xi) \right]$.
The Maxwellian plasma dispersion, $Z(\xi) = \frac{1}{\sqrt{\pi}} \int_{-\infty}^{\infty} \frac{e^{-z^2}}{z - \xi} dz$ is efficiently computed via the Pad\'{e} approximation method with the $J$-pole expansion \cite{Ronnmark_1983}, $Z(\xi) \approx Z_J(\xi) = \sum_{j=1}^J \frac{b_j}{\xi - c_j}$, where the coefficients $b_j$ and $c_j$ are tabulated by Xie et al. \cite{xie2016}. For subsequent derivations, we further employ the relations from \cite{xie2019} $\sum_{j} b_j = -1$, $\sum_{j} b_j c_j = 0$, and $\sum_{j} b_j c_j^2 = -1/2$. When both $\kappa_{\parallel s}\rightarrow \infty$ and $\kappa_{\perp s}\rightarrow \infty$, the coefficients in equation set (\ref{Cn_pbk_rational_expansion}) reduce to
\begin{eqnarray}
\label{Csn_max_loss-cone}
\begin{cases}
\begin{split}
\lim_{\kappa_{\parallel s},\kappa_{\perp s}\rightarrow \infty} C_{12sn}^\mathrm{PBK}
= \left(\frac{n \omega_{cs}}{k_{\parallel} \theta_{\parallel s}}\right) 
\tilde S_{17} Z(\xi_n) 
-\left(\frac{\theta_{\perp s}^2 }{2\theta_{\parallel s}^2}\right) 
\tilde S_2 Z^{\prime}(\xi_n)
= \sum_{j=1}^J \frac{\tilde b_{12snj}}{\omega-\tilde c_{snj}},
\\
\lim_{\kappa_{\parallel s},\kappa_{\perp s}\rightarrow \infty} C_{34sn}^\mathrm{PBK} 
= \left(\frac{n \omega_{cs}}{k_{\parallel} \theta_{\parallel s}} \right)
\tilde S_{38} Z(\xi_n) 
-\left(\frac{\theta_{\perp s}^2}{2\theta_{\parallel s}^2} \right)
\tilde S_4 Z^{\prime}(\xi_n)
= \sum_{j=1}^J \frac{\tilde b_{34snj}}{\omega-\tilde c_{snj}},
\\
\lim_{\kappa_{\parallel s},\kappa_{\perp s}\rightarrow \infty} C_{56sn}^\mathrm{PBK} 
= \left(\frac{n\omega_{cs}}{k_{\parallel} \theta_{\parallel s}}\right)
\tilde S_{59} Z(\xi_n)
-\left(\frac{\theta_{\perp s}^2}{2\theta_{\parallel s}^2}\right) 
\tilde S_6 Z^{\prime}(\xi_n)
= \sum_{j=1}^J \frac{\tilde b_{56snj}}{\omega-\tilde c_{snj}},
\end{split}
\end{cases}
\end{eqnarray}
where
\begin{eqnarray}
\label{bxsn_max_loss-cone}
\begin{cases}
\begin{split}
&\tilde c_{snj} = n\omega_{cs} + k_{\parallel}u_{s0} + c_j k_{\parallel} \theta_{\parallel s},
\\
&\tilde b_{12snj} = n\omega_{cs}b_j \tilde S_{17} +  b_j c_j k_{\parallel} \theta_{\parallel s} \frac{T_{\perp s}\tilde S_{2}}{T_{\parallel s}(\sigma_s + 1)},
\\
&\tilde b_{34snj} = n\omega_{cs}b_j \tilde S_{38} +  b_j c_j k_{\parallel} \theta_{\parallel s} \frac{T_{\perp s}\tilde S_{4}}{T_{\parallel s}(\sigma_s + 1)},
\\
&\tilde b_{56snj} = n\omega_{cs}b_j \tilde S_{59} +  b_j c_j k_{\parallel} \theta_{\parallel s} \frac{T_{\perp s}\tilde S_{6}}{T_{\parallel s}(\sigma_s + 1)}.
\end{split}
\end{cases}
\end{eqnarray}
Substituting the coefficients from equation set (\ref{Csn_max_loss-cone}) into the PBK susceptibility $\chi_{ij}^\mathrm{PBK}$ in equation set (\ref{chi_PBK}), yields all susceptibility components for the loss-cone drift BM plasma.

\section{Equivalent linear system for the BO-PBK solver}
\label{sec3}
The Vlasov-Maxwell system provides a complete kinetic description of particle-electromagnetic field interactions in plasmas. From its linearized equations,  one derives a general dispersion relation for linear wave modes. Within this self-consistent theoretical framework, we have developed a novel numerical solver for kinetic waves in a uniformly magnetized plasma under the small-amplitude perturbations. The solver models the plasma as a responsive medium characterized by a conductivity tensor derived from the linearized Vlasov-Maxwell equations. Integrating the tensor into Maxwell's equations via the Ohm's law yields a closed linear system for solving the plasma wave and instability modes.
The perturbation Maxwell's equations are
\begin{subequations}
\begin{align}
\label{Maxwell_Eq1}
\nabla \times \delta \boldsymbol{E} &= 
-\frac{\partial \delta \boldsymbol{B}}{\partial t},
\\
\label{Maxwell_Eq2}
\nabla \times \delta \boldsymbol{B} &=
\mu_0 \delta \boldsymbol{J} + \mu_0 \epsilon_0 \frac{\partial\delta \boldsymbol{E}}{\partial t} 
= \mu_0 \epsilon_0 \boldsymbol{\mathcal\epsilon} \cdot 
\frac{\partial \delta \boldsymbol{E}}{\partial t}.
\end{align}
\end{subequations}
The conductivity tensor $\boldsymbol{\mathcal\sigma} =  -i\epsilon_0 \omega\boldsymbol{\mathcal\chi}$ defines the dielectric tensor $\boldsymbol{\mathcal\epsilon} = \boldsymbol{\mathcal I} + \frac{i}{\epsilon_0 \omega}\boldsymbol{\mathcal\sigma}$. The current density response in the frequency domain is then
\begin{equation}
\label{OhmsLaw_matrix}
\begin{pmatrix}
\delta J_x \\ \delta J_y \\ \delta J_z 
\end{pmatrix}
= 
-i\epsilon_0 \omega 
\begin{pmatrix}
\chi_{11} & \chi_{12} & \chi_{13} \\
\chi_{21} & \chi_{22} & \chi_{23} \\
\chi_{31} & \chi_{32} & \chi_{33}
\end{pmatrix} 
\begin{pmatrix}
\delta E_x \\ \delta E_y \\ \delta E_z
\end{pmatrix},
\end{equation}
where $\chi_{ij} = \sum_{s=1}^{S_\mathrm{PBK}} \chi_{s,ij}^\mathrm{PBK} + \sum_{s=1}^{S_\mathrm{KM}} \chi_{s,ij}^\mathrm{KM}$. Here, $S_\mathrm{PBK}$ and $S_\mathrm{KM}$ denote the number of PBK and KM particle species, respectively, and $S = S_\mathrm{PBK} + S_\mathrm{KM}$ represents the total number of species. The PBK model generalizes the BM model, which is recovered as $\kappa \rightarrow \infty$ \cite{baiwei2023,Nazeer2020}.
To analyze electromagnetic waves propagation, we apply the Fourier ansatz $\partial/\partial t = -i\omega$ to Maxwell's equations (\ref{Maxwell_Eq1}), (\ref{Maxwell_Eq2}), yielding the dispersion relation $\det\left(\boldsymbol{k}\boldsymbol{k} - k^2 \boldsymbol{\mathcal I} + \frac{\omega^2}{c^2}\boldsymbol{\mathcal\epsilon}\right) = 0$. Instead of employing iterative root-finding algorithms (such as the shooting or M\"{u}ller method) with initial guesses for $\omega$, we reformulate the system of equations (\ref{Maxwell_Eq1}), (\ref{Maxwell_Eq2}), (\ref{OhmsLaw_matrix}) as a linear eigenvalue problem. This approach constructs a unified linear system, detailed in the following section.
\subsection{Equivalent linear system for PBK and KM plasmas}
The perturbed current in the $x$-direction, derived from Eqs. (\ref{Cn12sn_rational_expansion}), (\ref{Cn34sn_rational_expansion}), (\ref{ch11_PBK}),  (\ref{ch12_PBK}), (\ref{ch13_PBK}) and Ohm's law (\ref{OhmsLaw_matrix}), is
\begin{eqnarray}
\label{Jx_pbk}
\begin{split}
\delta J_{x}^\mathrm{PBK} =& \frac{b_{x10}}{\omega} \delta E_x  
+ \sum_{sn}\sum_{l=1}^{\kappa_{\parallel_s}} 
\frac{b_{x33sn,l+1}}{\omega(\omega - c_{sn})^{l}} \delta E_z
\\
&+\sum_{sn}\sum_{l=1}^{\kappa_{\parallel_s}+1} \left[
\frac{b_{x11snl}}{\omega \left(\omega - c_{sn}\right)^l} \delta E_x
+ \frac{b_{x21snl}}{\omega (\omega - c_{sn})^l}\delta E_y
+ \frac{b_{x31snl}}{\omega(\omega - c_{sn})^{l}} \delta E_z \right]
\\
&+ \sum_{sn}\sum_{l=1}^{\kappa_{\parallel_s}+1} \left[
\frac{b_{x12snl}}{\omega \left(\omega - c_{sn}\right)^{l+1}} \delta E_x
+ \frac{b_{x22snl}}{\omega (\omega - c_{sn})^{l+1}} \delta E_y
+ \frac{b_{x32snl}}{\omega(\omega - c_{sn})^{l+1}} \delta E_z \right],
\end{split}
\end{eqnarray}
where
\begin{eqnarray}
\label{bx_coef}
\begin{cases}
\begin{split}
&b_{x10} = i\epsilon_0 \sum_s \omega_{ps}^2,
\quad
b_{x30}  = -i\epsilon_0  \tan\theta\sum_s \omega_{ps}^2,
\quad
b_{x11snl} = -i\epsilon_0 \omega_{ps}^2 n^2  b_{1snl},
\\
&b_{x12snl} = -i\epsilon_0 \omega_{ps}^2 n^2 b_{2snl},
\quad
b_{x21snl} = \epsilon_0 \omega_{ps}^2 n b_{3snl},
\quad
b_{x22snl} = \epsilon_0 \omega_{ps}^2 n b_{4snl},
\\
&b_{x31snl} = -i\epsilon_0 \tan\theta \omega_{ps}^2 n
\left[\frac{(c_{sn}-n\omega_{cs})b_{1snl} + b_{2snl}}{\omega_{cs}} \right],
\\
&b_{x32snl} =  -i\epsilon_0 \tan\theta \omega_{ps}^2 n
\left[\frac{(c_{sn}-n\omega_{cs})b_{2snl}}{\omega_{cs}}\right],
\\
&b_{x33snl} = -i\epsilon_0 \tan\theta \omega_{ps}^2 n
\frac{b_{1snl}}{\omega_{cs}},
\quad
\sum_{sn} b_{x33sn1} = -b_{x30}.
\end{split}
\end{cases}
\end{eqnarray}
Furthermore, Eq. (\ref{Jx_pbk}) leads to the linear eigenvalue problem
$\omega \boldsymbol{x} = \boldsymbol{M}_{x} \boldsymbol{x}$. Expanding this system yields,
\begin{eqnarray}
\label{xsnlj_matrix_kappa}
\left\{
\begin{array}{c}
\begin{split}
&\omega \delta J_{x}^\mathrm{PBK} 
= b_{x10} \delta E_x + \sum_{sn}\sum_{l=1}^{\kappa_{\parallel_s}+1}x_{snl1},
\\
&\omega x_{snl1} = c_{sn} x_{snl1} + x_{snl2},
\\
&\qquad \quad \vdots\\
&\omega x_{snl,l-1} = c_{sn} x_{snl,l-1} + x_{snll}, 
\\
&\omega x_{snll} = 
\begin{cases}
&b_{x33sn,l+1} \delta E_z  + b_{x11snl}\delta E_x + b_{x21snl}\delta E_y 
+ b_{x31snl}\delta E_z 
\\
&+ c_{sn} x_{snll}+ x_{snl,l+1}, \text{(if $l \le \kappa_{\parallel s}$)}, 
\\
&b_{x11snl}\delta E_x + b_{x21snl}\delta E_y 
+ b_{x31snl}\delta E_z + c_{sn} x_{snll}+ x_{snl,l+1}, 
\text{(if $l = \kappa_{\parallel s}+1$)},
\end{cases}
\\
&\omega x_{snl,l+1} = b_{x12snl} \delta E_x + b_{x22snl} \delta E_y 
+ b_{x32snl} \delta E_z + c_{sn} x_{snl,l+1},
\end{split}
\end{array}
\right.
\end{eqnarray}
The vector is defined as $\boldsymbol{x} = \left[x_{snl1},x_{snl2},\cdots, x_{snl,l+1},\delta J_{x}^\mathrm{PBK}\right]^{T}$, and therefore the total number of equations in linear system  (\ref{xsnlj_matrix_kappa}) is $N_{x}^\mathrm{PBK} = 1+ \sum_{s=1}^{S_\mathrm{PBK}}\sum_{n=-N_s}^{N_s}\sum_{l=1}^{\kappa_{\parallel s}+1}\left(l+1\right)$.

Similarly, the perturbed current in the $y$-direction is given by
\begin{eqnarray}
\label{Jy_pbk}
\begin{split}
\delta J_{y}^\mathrm{PBK} 
=& \frac{b_{y20}}{\omega} \delta E_y + \sum_{sn}\sum_{l=1}^{\kappa_{\parallel s}} 
\frac{b_{y33sn,l+1}}{\omega \left(\omega-c_{sn}\right)^{l}} \delta E_z, 
\\
&+ \sum_{sn}\sum_{l=1}^{\kappa_{\parallel_s}+1} \left[
\frac{b_{y11snl}}{\omega (\omega - c_{sn})^l}\delta E_x
+ \frac{b_{y21snl}}{\omega \left(\omega - c_{sn}\right)^l} \delta E_y
+ \frac{b_{y31snl}}{\omega \left(\omega-c_{sn}\right)^{l}} \delta E_z \right]
\\
&+ \sum_{sn}\sum_{l=1}^{\kappa_{\parallel_s}+1} \left[
\frac{b_{y12snl}}{\omega (\omega - c_{sn})^{l+1}} \delta E_x
+ \frac{b_{y22snl}}{\omega \left(\omega - c_{sn}\right)^{l+1}} \delta E_y
+ \frac{b_{y32snl}}{\omega \left(\omega-c_{sn}\right)^{l+1}} \delta E_z \right],
\end{split}
\end{eqnarray}
as derived from Eqs. (\ref{Cn12sn_rational_expansion}),(\ref{Cn34sn_rational_expansion}), (\ref{Cn56sn_rational_expansion}),  (\ref{ch12_PBK}),  (\ref{ch22_PBK}), (\ref{ch23_PBK}), and (\ref{OhmsLaw_matrix}). The coefficients are
\begin{eqnarray}
\label{by_coef}
\begin{cases}
\begin{split}
&b_{y20} = i \epsilon_0 \sum_s \omega_{ps}^2,
\quad
b_{y11snl} = -\epsilon_0 \omega_{ps}^2 n b_{3snl},
\quad
b_{y12snl} = -\epsilon_0 \omega_{ps}^2 n b_{4snl},
\\
&b_{y21snl} = -i\epsilon_0 \omega_{ps}^2 b_{5snl},
\quad
b_{y22snl} = -i\epsilon_0 \omega_{ps}^2 b_{6snl},
\\
&b_{y31snl} = -\epsilon_0 \tan\theta \omega_{ps}^2 
\left[ \frac{\left(c_{sn} - n\omega_{cs}\right)b_{3snl} + b_{4snl}}{\omega_{cs}}  \right],
\\
&b_{y32snl} = -\epsilon_0 \tan\theta \omega_{ps}^2
\left[\frac{\left(c_{sn} - n\omega_{cs}\right) b_{4snl}}{\omega_{cs}} \right],
\\
&b_{y33snl} = -\epsilon_0  \tan\theta \omega_{ps}^2
\left(\frac{ b_{3snl}}{\omega_{cs}} \right),
\quad
\sum_{sn} b_{y33sn1} = 0.
\end{split}
\end{cases}
\end{eqnarray}
Applying the same method to Eq. (\ref{Jy_pbk}) yields the linear system for the $y$-direction $\omega \boldsymbol{y} = \boldsymbol{M}_{y} \boldsymbol{y}$, its explicit form is provided below
\begin{eqnarray}
\label{ysnlj_matrix_kappa}
\left\{
\begin{array}{c}
\begin{split}
&\omega \delta J_{y}^\mathrm{PBK} 
= b_{y20} \delta E_y + \sum_{sn}\sum_{l=1}^{\kappa_{\parallel_s}+1}  y_{snl1}, 
\\
&\omega y_{snl1} = c_{sn} y_{snl1}  + y_{snl2},
\\
&\qquad \quad \vdots \\
&\omega y_{snl,l-1} = c_{sn} y_{snl,l-1}  + y_{snll},
\\
&\omega y_{snll} = 
\begin{cases}
&b_{y33sn,l+1}\delta E_z +  b_{y11snl}\delta E_x + b_{y21snl}\delta E_y 
+ b_{y31snl}\delta E_z 
\\
&+ c_{sn} y_{snll}  + y_{snl,l+1}, \text{(if $l\le \kappa_{\parallel s}$)},
\\
&b_{y11snl}\delta E_x + b_{y21snl}\delta E_y 
+ b_{y31snl}\delta E_z + c_{sn} y_{snll}  + y_{snl,l+1}, 
\text{(if $l= \kappa_{\parallel s}+1$)},
\end{cases}
\\
&\omega y_{snl,l+1} = b_{y12snl} \delta E_x + b_{y22snl} \delta E_y 
+ b_{y32snl} \delta E_z + c_{sn} y_{snl,l+1},
\end{split}
\end{array}
\right.
\end{eqnarray}
where $\boldsymbol{y} = \left[y_{snl1},y_{snl2},\cdots, y_{snl,l+1},\delta J_{y}^\mathrm{PBK}\right]^{T}$, the total number of equations in linear system  (\ref{ysnlj_matrix_kappa}) is $N_{y}^\mathrm{PBK} = N_{x}^\mathrm{PBK}$.

Similarly, the perturbed current in the $z$-direction is given by
\begin{eqnarray}
\label{Jz_pbk}
\begin{split}
\delta J_{z}^\mathrm{PBK} =& i\epsilon_0 \sum_s \omega_{ps} \delta E_z
+ \sum_{sn}\sum_{l=1}^{\kappa_{\parallel s}-1} 
\frac{b_{z34sn,l+2}}{\omega \left(\omega - c_{sn}\right)^{l}} \delta E_z, 
\\
&+ \sum_{sn}\sum_{l=1}^{\kappa_{\parallel s}} \left[
\frac{b_{z13sn,l+1}}{\omega \left(\omega-c_{sn}\right)^{l}} \delta E_x
+ \frac{b_{z23sn,l+1}}{\omega \left(\omega-c_{sn}\right)^{l}} \delta E_y
+ \frac{b_{z33sn,l+1}}{\omega \left(\omega - c_{sn}\right)^{l}} \delta E_z \right]
\\
&+ \sum_{sn}\sum_{l=1}^{\kappa_{\parallel s}+1} \left[
\frac{b_{z11snl}}{\omega \left(\omega-c_{sn}\right)^l} \delta E_x
+ \frac{b_{z21snl}}{\omega \left(\omega-c_{sn}\right)^{l}} \delta E_y
+ \frac{b_{z31snl}}{\omega \left(\omega - c_{sn}\right)^{l}} \delta E_z \right]
\\
&+ \sum_{sn}\sum_{l=1}^{\kappa_{\parallel s}+1} \left[
\frac{b_{z12snl}}{\omega \left(\omega-c_{sn}\right)^{l+1}} \delta E_x
+ \frac{b_{z22snl}}{\omega \left(\omega-c_{sn}\right)^{l+1}} \delta E_y
+ \frac{b_{z32snl}}{\omega \left(\omega - c_{sn}\right)^{l+1}} \delta E_z \right],
\end{split}
\end{eqnarray}
where
\begin{eqnarray}
\label{bz_coef}
\begin{cases}
\begin{split}
&b_{z10} = -i\epsilon_0 \tan\theta \sum_s \omega_{ps}^2,
\quad
b_{z11snl} =  -i\epsilon_0 \tan\theta \omega_{ps}^2 n
\left[\frac{\left(c_{sn} - n\omega_{cs}\right)b_{1snl} + b_{2snl}}{\omega_{cs}}\right],
\\
&b_{z30} = i\epsilon_0 \tan^2\theta \sum_s \omega_{ps}^2,
\quad
b_{z12snl} = -i\epsilon_0 \tan\theta \omega_{ps}^2 n
\left[\frac{\left(c_{sn} - n\omega_{cs}\right)b_{2snl}}{\omega_{cs}}\right],
\\
&b_{z13snl} = -i\epsilon_0 \tan\theta \omega_{ps}^2 n
\left(\frac{b_{1snl}}{\omega_{cs}} \right),
\quad
b_{z21snl} = \epsilon_0 \tan\theta \omega_{ps}^2
\left[ \frac{\left(c_{sn} - n\omega_{cs}\right)b_{3snl} + b_{4snl}}{\omega_{cs}}\right],
\\
&b_{z22snl} = \epsilon_0 \tan\theta \omega_{ps}^2
\left[\frac{\left(c_{sn} - n\omega_{cs}\right) b_{4snl}}{\omega_{cs}} \right],
\quad
b_{z23snl} = \epsilon_0 \tan\theta \omega_{ps}^2
\left(\frac{ b_{3snl}}{\omega_{cs}} \right),
\\
&b_{z31snl} = -i\epsilon_0 \tan^2\theta \omega_{ps}^2
\left[\frac{\left(c_{sn} - n\omega_{cs}\right)^2 b_{1snl} + 2\left(c_{sn} - n\omega_{cs}\right) b_{2snl}}{\omega_{cs}^2} \right],
\\
&b_{z32snl} = -i\epsilon_0 \tan^2\theta \omega_{ps}^2
\left[\frac{\left(c_{sn} - n\omega_{cs}\right)^2 b_{2snl}}{\omega_{cs}^2} \right],
\\
&b_{z33snl} = -i\epsilon_0 \tan^2\theta \omega_{ps}^2
\left[\frac{2\left(c_{sn} - n\omega_{cs}\right) b_{1snl} + b_{2snl}}{\omega_{cs}^2} \right],
\quad
b_{z34snl} = -i\epsilon_0 \tan^2\theta \omega_{ps}^2
\left(\frac{b_{1snl}}{\omega_{cs}^2}\right),
\\
&\sum_{sn}b_{z13sn1} = -b_{z10},
\quad
\sum_{sn}b_{z23sn1} = 0,
\quad
\sum_{sn}b_{z33sn1} = i\epsilon_0 \sum_{s} \omega_{ps}^2,
\\
&\sum_{sn} c_{sn} b_{z34sn1} = b_{z30},
\quad
\sum_{sn} b_{z34sn2} = 0,
\quad
\sum_{sn} b_{z34sn1} = 0.
\end{split}
\end{cases}
\end{eqnarray}
Following the same methodology, Eq. (\ref{Jz_pbk}) for $z$-direction yields the linear system $\omega \boldsymbol{z} = \boldsymbol{M}_{z} \boldsymbol{z}$, the explicit form is given by
\begin{eqnarray}
\label{zsnlj_matrix_kappa}
\left\{
\begin{array}{c}
\begin{split}
&\omega \delta J_{z}^\mathrm{PBK} 
= i\epsilon_0 \sum_{s} \omega_{ps}^2 \delta E_z  
+ \sum_{sn}\sum_{l=1}^{\kappa_{\parallel s}+1}z_{snl1}, 
\\
&\omega z_{snl1}  = c_{sn} z_{snl1} + z_{snl2},
\\
&\qquad \quad \vdots \\
&\omega z_{snl,l-1}  = c_{sn} z_{snl,l-1} + z_{snll},
\\
&\omega z_{snll}=
\begin{cases}
& b_{z34sn,l+2} \delta E_z  +  b_{z13sn,l+1} \delta E_x + b_{z23sn,l+1} \delta E_y + b_{z33sn,l+1} \delta E_z
\\
&+ b_{z11snl}\delta E_x + b_{z21snl}\delta E_y +  b_{z31snl}\delta E_z
+ c_{sn} z_{snll} + z_{snl,l+1}, \text{(if $l\le \kappa_{\parallel s}-1$)},
\\
& b_{z13sn,l+1} \delta E_x + b_{z23sn,l+1} \delta E_y + b_{z33sn,l+1} \delta E_z
\\
&+ b_{z11snl}\delta E_x + b_{z21snl}\delta E_y +  b_{z31snl}\delta E_z
+ c_{sn} z_{snll} + z_{snl,l+1}, \text{(if $l= \kappa_{\parallel s}$)},
\\
&b_{z11snl}\delta E_x + b_{z21snl}\delta E_y +  b_{z31snl}\delta E_z
+ c_{sn} z_{snll} + z_{snl,l+1}, \text{(if $l= \kappa_{\parallel s}+1$)},
\end{cases}
\\
&\omega z_{snl,l+1} = b_{z12snl} \delta E_x + b_{z22snl} \delta E_y + b_{z32snl} \delta E_z + c_{sn} z_{snl,l+1},
\end{split}
\end{array}
\right.
\end{eqnarray}
where $\boldsymbol{z} = \left[z_{snl1},z_{snl2},\cdots, z_{snl,l+1},\delta J_{z}^\mathrm{PBK}\right]^T$, the total number of equations in linear system (\ref{zsnlj_matrix_kappa}) is $N_{z}^\mathrm{PBK} = N_{x}^\mathrm{PBK}$.

The above derivation constructs the total linear system for PBK plasma by combining the component systems in the $x$-, $y$-, and $z$-directions, given by Eqs. (\ref{xsnlj_matrix_kappa}), (\ref{ysnlj_matrix_kappa}), and (\ref{zsnlj_matrix_kappa}), respectively. The total dimension of this system is $N_\mathrm{PBK} = 3 N_{x}^\mathrm{PBK}$. 
Furthermore, Eqs. (\ref{proof_jx}) and (\ref{proof_jz}) establish the equivalence between the component systems for the $x$- and $z$-directions (\ref{xsnlj_matrix_kappa}) and (\ref{zsnlj_matrix_kappa}), and their respective perturbed currents, (\ref{Jx_pbk}) and (\ref{Jz_pbk}).

\subsection{Equivalent linear system for BM plasmas}
The perturbed current in BM plasma is derived from equations (\ref{Csn_max_loss-cone}) and (\ref{chi_PBK}) combined with Ohm's law (\ref{OhmsLaw_matrix}), giving the simplified expressions \cite{bai2024,xie2016},
\begin{equation}
\label{Jxyz_BM}
\begin{pmatrix}  
\delta J_x^\mathrm{BM} \\
 \delta J_y^\mathrm{BM}  \\
 \delta J_z^\mathrm{BM} 
\end{pmatrix}
=
\begin{pmatrix} 
\frac{b_{x1}^\mathrm{BM} }{\omega} + \sum_{snj} \frac{b_{x1snj}^\mathrm{BM}}{\omega-\tilde{c}_{snj}} 
& \frac{b_{x2}^\mathrm{BM}}{\omega} + \sum_{snj} \frac{b_{x2snj}^\mathrm{BM}}{\omega - \tilde{c}_{snj}}
& \frac{b_{x3}^\mathrm{BM}}{\omega} + \sum_{snj} \frac{b_{x3snj}^\mathrm{BM}}{\omega - \tilde{c}_{snj}}
\\
\frac{b_{y1}^\mathrm{BM}}{\omega} + \sum_{snj}\frac{b_{y1snj}^\mathrm{BM}}{\omega - \tilde{c}_{snj}}  
& \frac{b_{y2}^\mathrm{BM}}{\omega} + \sum_{snj} \frac{b_{y2snj}^\mathrm{BM}}{\omega-\tilde{c}_{snj}}
& \frac{b_{y3}^\mathrm{BM}}{\omega} + \sum_{snj} \frac{b_{y3snj}^\mathrm{BM}}{\omega - \tilde{c}_{snj}} 
\\
\frac{b_{z1}^\mathrm{BM}}{\omega} + \sum_{snj} \frac{b_{z1snj}^\mathrm{BM}}{\omega - \tilde{c}_{snj}}
& \frac{b_{z2}^\mathrm{BM}}{\omega} + \sum_{snj} \frac{b_{z2snj}^\mathrm{BM}}{\omega - \tilde{c}_{snj}} 
& \frac{b_{z3}^\mathrm{BM}}{\omega} + \sum_{snj} \frac{b_{z3snj}^\mathrm{BM}}{\omega - \tilde{c}_{snj}} 
\end{pmatrix}
\begin{pmatrix}  
\delta E_x  \\
 \delta E_y  \\
 \delta E_z
\end{pmatrix},
\end{equation}
where
\begin{eqnarray}
\begin{cases}
b_{x1}^\mathrm{BM} 
= i\epsilon_0 \left(\sum_s \omega_{ps}^2
+ \sum_{snj}\omega_{ps}^2 n^2 \frac{\tilde{b}_{12snj}}{\tilde{c}_{snj}} \right),
\quad
b_{x2}^\mathrm{BM} 
= -\epsilon_0 \sum_{snj} \omega_{ps}^2 n \frac{\tilde{b}_{34snj}}{\tilde{c}_{snj}},
\\
b_{x3}^\mathrm{BM} 
= -i\epsilon_0 \tan\theta \left( \sum_{s} \omega_{ps}^2
+ \sum_{snj} \omega_{ps}^2 n^2 \frac{\tilde{b}_{12snj}}{\tilde{c}_{snj}} \right),
\quad
b_{x1snj}^\mathrm{BM} 
= -i\epsilon_0 \omega_{ps}^2 n^2 \frac{\tilde{b}_{12snj}}{\tilde{c}_{snj}},
\\
b_{x2snj}^\mathrm{BM} 
= \epsilon_0 \omega_{ps}^2 n \frac{\tilde{b}_{34snj}}{\tilde{c}_{snj}},
\quad
b_{x3snj}^\mathrm{BM} 
= -i\epsilon_0 \tan\theta \omega_{ps}^2 n
\frac{\left(1-n\omega_{cs}/\tilde{c}_{snj} \right)\tilde{b}_{12snj}}{\omega_{cs}},
\\
b_{y1}^\mathrm{BM} 
= \epsilon_0 \sum_{snj}\omega_{ps}^2 n \frac{\tilde{b}_{34snj}}{\tilde{c}_{snj}},
\quad
b_{y2}^\mathrm{BM} 
= i\epsilon_0 \left( \sum_s \omega_{ps}^2 + \sum_{snj}\omega_{ps}^2
\frac{\tilde{b}_{56snj}}{\tilde{c}_{snj}} \right),
\\
b_{y3}^\mathrm{BM} 
= -\epsilon_0 \tan\theta  \sum_{snj} \omega_{ps}^2 n 
\frac{\tilde{b}_{34snj}}{\tilde{c}_{snj}},
\quad
b_{y1snj}^\mathrm{BM} 
= -\epsilon_0 \omega_{ps}^2 n \frac{\tilde{b}_{34snj}}{\tilde{c}_{snj}},
\\
b_{y2snj}^\mathrm{BM} 
= -i\epsilon_0 \omega_{ps}^2 \frac{\tilde{b}_{56snj}}{\tilde{c}_{snj}},
\quad
b_{y3snj}^\mathrm{BM} 
= -\epsilon_0 \tan\theta \omega_{ps}^2
\frac{\left(1 - n\omega_{cs}/\tilde{c}_{snj}\right)\tilde{b}_{34snj}}{\omega_{cs}},
\\
b_{z1}^\mathrm{BM}
= -i\epsilon_0 \tan\theta 
\left(\sum_{s} \omega_{ps}^2 + \sum_{snj} \omega_{ps}^2 n^2  
\frac{\tilde{b}_{12snj}}{\tilde{c}_{snj}} \right),
\quad
b_{z2}^\mathrm{BM} = \epsilon_0 \tan\theta \sum_{snj} \omega_{ps}^2 n
\frac{\tilde{b}_{34snj}}{\tilde{c}_{snj}},
\\
b_{z3}^\mathrm{BM} 
= i\epsilon_0 \tan^2\theta  
\left(\sum_s \omega_{ps}^2 + \sum_{snj} \omega_{ps}^2 n^2  
\frac{\tilde{b}_{12snj}}{\tilde{c}_{snj}}\right),
\quad
b_{z1snj}^\mathrm{BM}
= -i\epsilon_0 \tan\theta \omega_{ps}^2 n  
\frac{\left(1-n\omega_{cs}/\tilde{c}_{snj} \right)\tilde{b}_{12snj}}{\omega_{cs}},
\\
b_{z2snj}^\mathrm{BM}  
= \epsilon_0 \tan\theta \omega_{ps}^2
\frac{\left(1 - n\omega_{cs}/\tilde{c}_{snj}\right)\tilde{b}_{34snj}}{\omega_{cs}},
\quad
b_{z3snj}^\mathrm{BM} 
= -i\epsilon_0 \tan^2\theta  \omega_{ps}^2 \tilde{b}_{12snj}
\left(\frac{\tilde{c}_{snj}}{\omega_{cs}^2}- \frac{2n}{\omega_{cs}} + \frac{n^2}{\tilde{c}_{snj}} \right).
\end{cases}
\end{eqnarray}
The three perturbed currents in Eq. (\ref{Jxyz_BM}) form a linear subsystem given by
\begin{eqnarray}
\label{xsnj_maxwell}
\begin{cases}
\begin{split}
\omega \delta J_{x1}^\mathrm{BM} 
&= b_{x1}^\mathrm{BM} \delta E_x + b_{x2}^\mathrm{BM} \delta E_y + b_{x3}^\mathrm{BM} \delta E_z,
\\
\omega \delta J_{x2}^\mathrm{BM}  
&= \omega \sum_{snj} \tilde{x}_{snj} 
=  \sum_{snj} \tilde{c}_{snj} \tilde{x}_{snj} + \sum_{snj} b_{x1snj}^\mathrm{BM}\delta E_x 
+ \sum_{snj} b_{x2snj}^\mathrm{BM}\delta E_y + \sum_{snj} b_{x3snj}^\mathrm{BM}\delta E_z,
\\
\omega \tilde{x}_{snj}  &= \tilde{c}_{snj} \tilde{x}_{snj}
+ b_{x1snj}^\mathrm{BM}\delta E_x + b_{x2snj}^\mathrm{BM} \delta E_y + b_{x3snj}^\mathrm{BM}\delta E_z, 
\end{split}
\end{cases}
\end{eqnarray}
\begin{eqnarray}
\label{ysnj_maxwell}
\begin{cases}
\begin{split}
\omega \delta J_{y1}^\mathrm{BM} 
&= b_{y1}^\mathrm{BM} \delta E_x + b_{y2}^\mathrm{BM} \delta E_y + b_{y3}^\mathrm{BM} \delta E_z,
\\
\omega \delta J_{y2}^\mathrm{BM}  
&= \omega \sum_{snj} \tilde{y}_{snj} 
=  \sum_{snj} \tilde{c}_{snj} \tilde{y}_{snj} + \sum_{snj} b_{y1snj}^\mathrm{BM}\delta E_x 
+ \sum_{snj} b_{y2snj}^\mathrm{BM}\delta E_y + \sum_{snj} b_{y3snj}^\mathrm{BM}\delta E_z,
\\
\omega \tilde{y}_{snj}  &= \tilde{c}_{snj} \tilde{y}_{snj}
+ b_{y1snj}^\mathrm{BM}\delta E_x + b_{y2snj}^\mathrm{BM} \delta E_y + b_{y3snj}^\mathrm{BM}\delta E_z, 
\end{split}
\end{cases}
\end{eqnarray}
and
\begin{eqnarray}
\label{zsnj_maxwell}
\begin{cases}
\begin{split}
\omega \delta J_{z1}^\mathrm{BM} 
&= b_{z1}^\mathrm{BM} \delta E_x + b_{z2}^\mathrm{BM} \delta E_y + b_{z3}^\mathrm{BM} \delta E_z, 
\\
\omega \delta J_{z2}^\mathrm{BM}  
&= \omega \sum_{snj} \tilde{z}_{snj} 
=  \sum_{snj} \tilde{c}_{snj} \tilde{z}_{snj} + \sum_{snj} b_{z1snj}^\mathrm{BM}\delta E_x 
+ \sum_{snj} b_{z2snj}^\mathrm{BM}\delta E_y 
+ \sum_{snj} b_{z3snj}^\mathrm{BM}\delta E_z,
\\
\omega \tilde{z}_{snj}  &= \tilde{c}_{snj} \tilde{z}_{snj}
+ b_{z1snj}^\mathrm{BM}\delta E_x + b_{z2snj}^\mathrm{BM} \delta E_y + b_{z3snj}^\mathrm{BM}\delta E_z.
\end{split}
\end{cases}
\end{eqnarray}
Here, the perturbed currents are defined as $\delta J_{x}^\mathrm{BM} =  \delta J_{x1}^\mathrm{BM} + \delta J_{x2}^\mathrm{BM}$,  $\delta J_{y}^\mathrm{BM} = \delta J_{y1}^\mathrm{BM} + \delta J_{y2}^\mathrm{BM}$, $\delta J_{z}^\mathrm{BM} = \delta J_{z1}^\mathrm{BM} + \delta J_{z2}^\mathrm{BM}$, with the summation operator $\sum_{snj} = \sum_{s=1}^{S_\mathrm{BM}}\sum_{n=-N_s}^{N_s}\sum_{j=1}^J$. Each eigenvalue subsystem, for the $x$-, $y$-, or $z$-direction, consists of the corresponding equations (\ref{xsnj_maxwell}), (\ref{ysnj_maxwell}), and (\ref{zsnj_maxwell})  and contains of $N_{x}^\mathrm{BM} = N_{y}^\mathrm{BM} = N_{z}^\mathrm{BM} = 2+\sum_{1}^{S_\mathrm{BM}}\sum_{n=-N_s}^{N_s}J$ equations. Consequently, the total dimension of the linear system for the BM plasma is $N_\mathrm{BM} = 3N_{x}^\mathrm{BM}$.

\subsection{Equivalent closed linear system}
The complete PBK-BM linear system combines the PBK system (Eqs. (\ref{xsnlj_matrix_kappa}), (\ref{ysnlj_matrix_kappa}), and (\ref{zsnlj_matrix_kappa})), the BM system (Eqs. (\ref{xsnj_maxwell}), (\ref{ysnj_maxwell}), and (\ref{zsnj_maxwell})), and Maxwell's equations in their perturbed component form
\begin{subequations}
\label{Maxwell_xyz}
\begin{align}
\label{eq: Maxwell_Ex}
\omega \delta E_x &= c^2 k_z \delta B_y - i \frac{\delta J_x}{\epsilon_0}, 
\\
\label{eq: Maxwell_Ey}
\omega \delta E_y &= -c^2 k_z \delta B_x + c^2 k_x \delta B_z - i \frac{\delta J_y}{\epsilon_0}, 
\\
\label{eq: Maxwell_Ez}
\omega \delta E_z &= -c^2 k_x \delta B_y - i \frac{\delta J_z}{\epsilon_0},
\\
\label{eq: Maxwell_Bx}
\omega \delta B_x &= -k_z \delta E_y,
\\
\label{eq: Maxwell_By}
\omega \delta B_y &= k_z \delta E_x - k_x \delta E_z,
\\
\label{eq: Maxwell_Bz}
\omega \delta B_z &= k_x \delta E_y.
\end{align}
\end{subequations}
The perturbed currents in Eqs. (\ref{eq: Maxwell_Ex}), (\ref{eq: Maxwell_Ey}), and (\ref{eq: Maxwell_Ez}) arise from both the PBK and BM plasmas, and are decomposed into distinct components as follows: $\delta J_x = \delta J_x^\mathrm{PBK} + \delta J_{x1}^\mathrm{BM}+\delta J_{x2}^\mathrm{BM}$, $\delta J_y= \delta J_y^\mathrm{PBK} + \delta J_{y1}^\mathrm{BM} + \delta J_{y2}^\mathrm{BM}$, and $\delta J_z= \delta J_z^\mathrm{PBK}+ \delta J_{z1}^\mathrm{BM} + \delta J_{z2}^\mathrm{BM}$. 
The resulting closed PBK-BM linear system governing oblique electromagnetic wave propagation is given by $\omega \mathbf{x} = \boldsymbol{B} \mathbf{x}$. The vector is $\boldsymbol{\mathrm{x}} = \left[\boldsymbol{x}_\mathrm{PBK}, \boldsymbol{x}_\mathrm{BM}, \delta \boldsymbol{J}_{1}^\mathrm{BM}, \delta \boldsymbol{E}, \delta \boldsymbol{B} \right]^T$, and the matrix \(\boldsymbol{B}\) is defined as follows
\begin{equation}
\label{ClosedLinearSystem}
\boldsymbol{B}
= \left[
\boldsymbol{B}_\mathrm{PBK},
\boldsymbol{B}_\mathrm{BM},
\boldsymbol{B}_\mathrm{Maxwell}
\right]^T,
\end{equation}
where 
\begin{subequations}
\begin{align}
& \boldsymbol{x}_\mathrm{PBK} = \left[
\underbrace{x_{snl1},\cdots, x_{snl,l+1}, \delta J_{x}^\mathrm{PBK}}_{N_x^\mathrm{PBK}},
\cdots, 
\underbrace{z_{snl1},\cdots, z_{snl,l+1}, \delta J_{z}^\mathrm{PBK}}_{N_z^\mathrm{PBK}}
\right],
\\ 
& \boldsymbol{x}_\mathrm{BM}  = \left[
\underbrace{\tilde{x}_{snj}, \delta J_{x2}^\mathrm{BM}}_{N_x^\mathrm{BM}},
\cdots, 
\underbrace{\tilde{z}_{snj}, \delta J_{z2}^\mathrm{BM}}_{N_z^\mathrm{BM}}
\right],
\quad
\delta \boldsymbol{J}_{1}^\mathrm{BM} 
= \left[\delta J_{x1}^\mathrm{BM}, \delta J_{y1}^\mathrm{BM}, \delta J_{z1}^\mathrm{BM} \right],
\\
&\delta \boldsymbol{E} = \left[\delta E_x, \delta E_y, \delta E_z \right],
\quad
\delta \boldsymbol{B} = \left[\delta B_x, \delta B_y, \delta B_z \right]. 
\end{align}
\end{subequations}
The complete linear system for the plasma comprising PBK and BM components is represented by a sparse matrix of order $N^{\text{closed}}$, where $N^{\text{closed}} = N_{\mathrm{PBK}} + N_{\mathrm{BM}} + 6$ denotes the total number of degrees of freedom. This global system matrix $\boldsymbol{B}$ is assembled from the following submatrices: the PBK plasma submatrix $\boldsymbol{B}_{\mathrm{PBK}}$, of dimension $N_{\mathrm{PBK}} \times N^{\text{closed}}$, defined by Eqs. (\ref{xsnlj_matrix_kappa}), (\ref{ysnlj_matrix_kappa}), and (\ref{zsnlj_matrix_kappa}); the BM plasma submatrix $\boldsymbol{B}_{\mathrm{BM}}$, of dimension $N_{\mathrm{BM}} \times N^{\text{closed}}$, defined by Eqs. (\ref{xsnj_maxwell}), (\ref{ysnj_maxwell}), and (\ref{zsnj_maxwell}); and the Maxwell submatrix $\boldsymbol{B}_{\mathrm{Maxwell}}$, of dimension $6 \times N^{\text{closed}}$, determined by the Maxwell component equation set (\ref{Maxwell_xyz}).
For the independent PBK or BM plasma models, the corresponding closed linear systems are described by sparse matrices of orders $N_{\mathrm{PBK}}^{\text{closed}} = N_{\mathrm{PBK}} + 6$ and $N_{\mathrm{BM}}^{\text{closed}} = N_{\mathrm{BM}} + 6$, respectively. By setting $\kappa_{\perp s}$ to a sufficiently large value (e.g., 200) in the BO-PBK solver, we force the PBK plasma model to converge to the KM plasma, which numerically represents the Maxwellian limit ($\kappa_{\perp s}\rightarrow \infty$).  The computational efficiency of the BO-PBK solver is highly sensitive to $\kappa_{\parallel s}$ index, as high values cause a prohibitive increase in matrix dimension. For instance, in the BO-PBK solver with $N=10$ and $S=1$, increasing $\kappa_{\parallel}$ from $2$ to $14$ raises $N_\mathrm{PBK}^\text{closed}$ from $576$ to $8514$, while the resulting matrix dimensions remain only about one-half to one-third of those required by the BO-KM solver \cite{bai2025}. When $\kappa_{\parallel s}$ is large, the distribution approaches a Maxwellian. For computational convenience, the algorithm switches to the BM matrix for the $s$th species eigenvalue problem whenever $\kappa_{\parallel s}$ exceeds the threshold $\kappa_{\parallel s, th}$.

\section{Benchmarks and applications}
\label{sec4}
In this section, we report the results of applying the BO-PBK solver to several representative test cases to validate its performance. These cases were computed on a system equipped with an Intel Core i7-8850H CPU, using MATLAB's eigenvalue functions eig() or eigs() for solution.

\subsection*{Case 1: R-, L-, and P-mode waves}
\begin{figure}[t]
\centering
\includegraphics[scale=0.6]{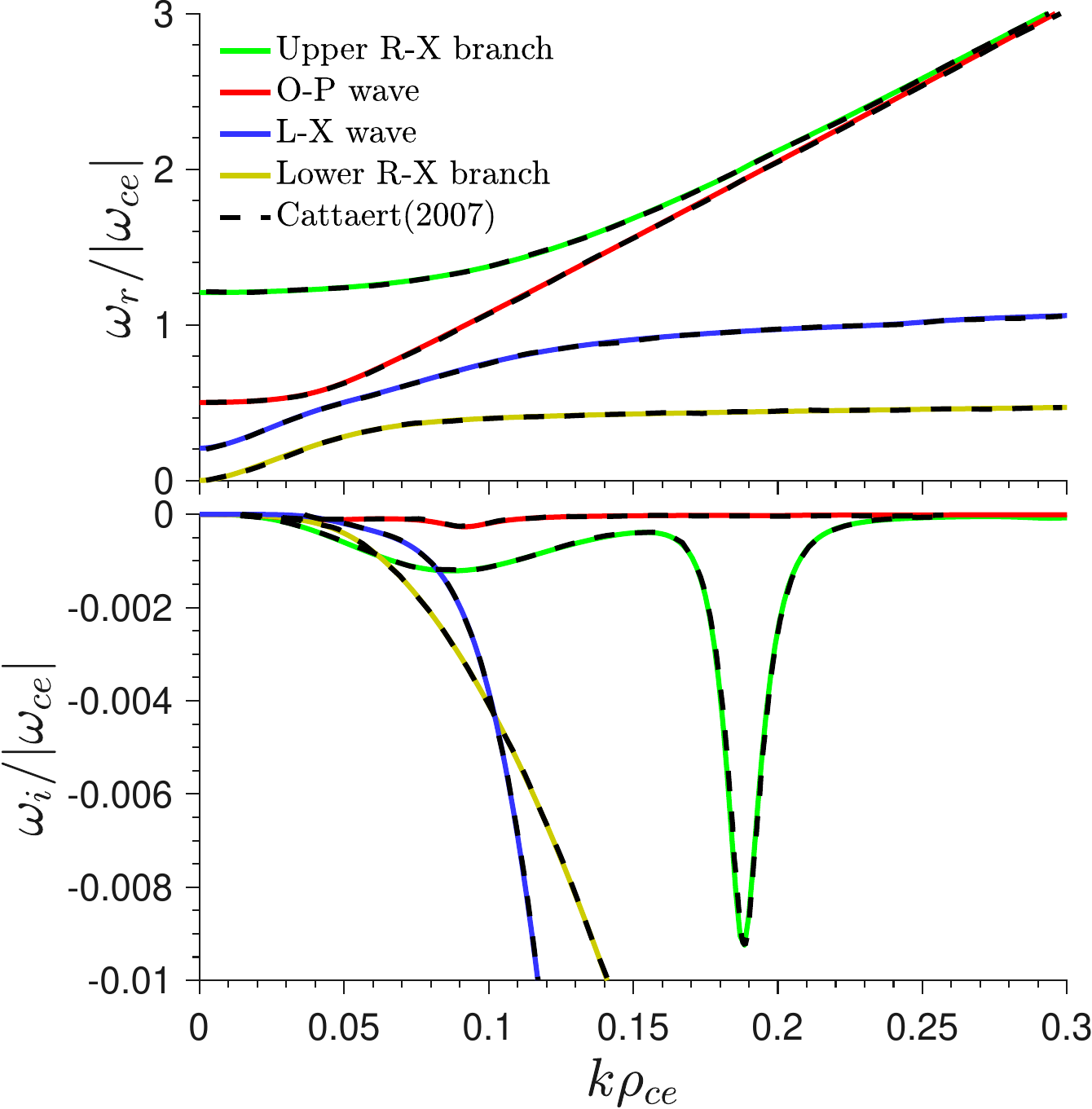}
\caption{$\omega_r/|\omega_{ce}|$ vs. $\rho_{ce}$ (top) and $\omega_i/|\omega_{ce}|$ vs. $\rho_{ce}$ (bottom) for $\kappa_{\parallel e}=1$, $\kappa_{\perp e}=200$, $\theta=30^\circ$, $\tilde\theta_{\perp e}/c=0.1$, $\omega_{pe}/|\omega_{ce}|=0.5$, $T_{\perp e}/T_{\parallel e}=1$, and $B_0=1.0\times 10^{-6}$ T, comparison with data from Fig. 1 of Cattaert et al. \cite{Cattaert2007} (black dashed lines).}
\label{benchmark_Cattaert2007Fig1}
\end{figure}

On the electron time scale, four modes are identified within the $\omega_{ce}$ range: the off-parallel L-X wave (left-hand circularly polarized), the upper and lower branches of the R-X wave (right-hand circularly polarized), and the O-P wave (ordinary plasma wave). At a fixed angle of $\theta = 30^{\circ}$, our numerical results (Fig. \ref{benchmark_Cattaert2007Fig1}) show excellent agreement with Ref. \cite{Cattaert2007}.

\subsection*{Case 2: Whistler instability}
\begin{figure}[t]
\centering
\includegraphics[scale=0.6]{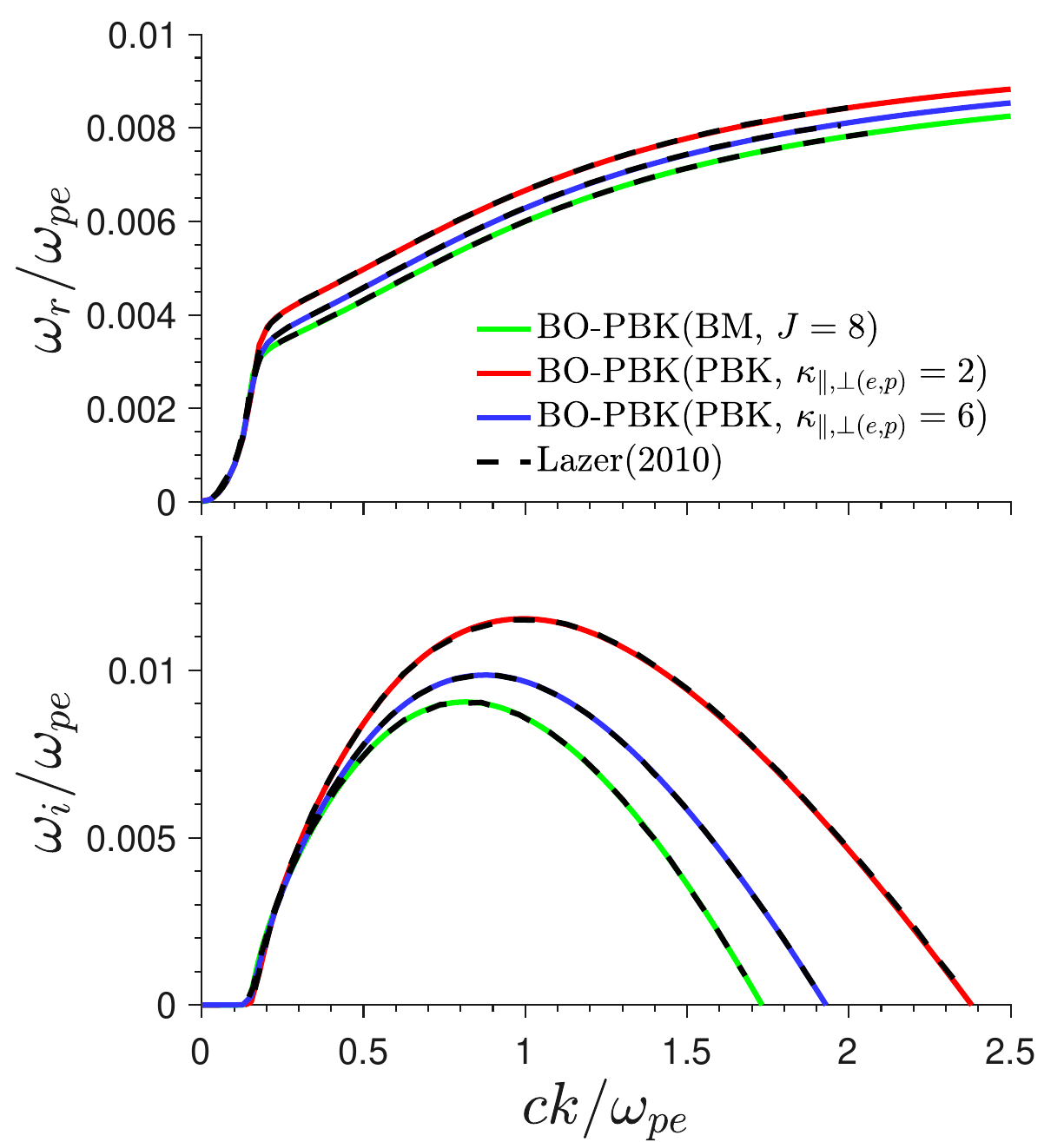}
\caption{
Comparison of real frequency $\omega_r$ (top) and growth rate $\omega_i$ (bottom) of parallel electron whistler-cyclotron modes for $\kappa_{\parallel,\perp(e,p)}=2,6,\infty$ in weakly magnetized plasmas ($|\omega_{ce}|/\omega_{pe} = 0.01$), with $v_{T\parallel e} = 0.02c$ and $T_{\perp(e,p)}/T_{\parallel(e,p)} = 4$. Data from Fig. 2 of Lazer et al. \cite{Lazar_Poedts2010}, are shown as black dashed lines.
}
\label{benchmark_lazar2010Fig2}
\end{figure}

The electron whistler instability is highly sensitive to anisotropic distribution shapes, making it ideal for BO-PBK solver validation. We compare with the weakly magnetized plasma ($|\omega_{ce}|/\omega_{pe} = 0.01$) studied by Lazar et al. \cite{Lazar_Poedts2010}, which consists electrons (e) and protons (p) with $T_{\perp(e,p)}/T_{\parallel(e,p)} = 4$. Figure \ref{benchmark_lazar2010Fig2} shows the real frequencies and growth rates of parallel electron whistler-cyclotron modes for PBK plasma ($\kappa_{\parallel, \perp(e,p)} = 2,6$) and BM plasma. The BO-PBK results (colored solid lines) agree well with the reference data (black dashed lines) from \cite{Lazar_Poedts2010}.

\begin{figure}[t]
\centering
\includegraphics[scale=0.6]{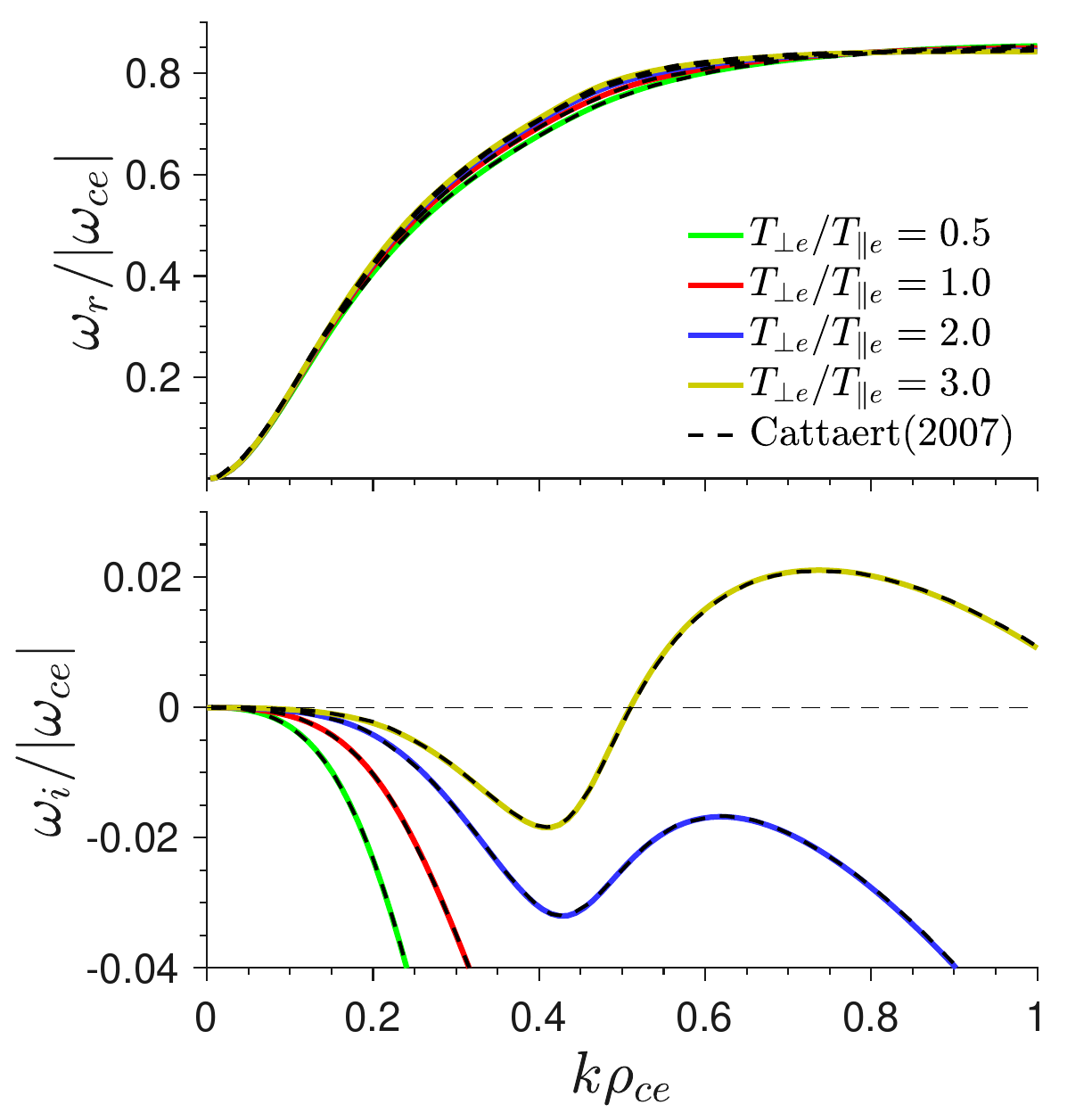}
\caption{Comparison of whistler wave real frequency $\omega_r$ vs. $k$ (top)  and growth rate $\omega_i$ vs. $k$ (bottom) with varying temperature anisotropy. Parameters: $\kappa_{\parallel e} = 1$, $\kappa_{\perp e} = 200$, $\theta = 30^{\circ}$, $\theta_{\perp e}/c = 0.1$, $|\omega_{ce}|/\omega_{pe} = 0.5$, and $B_0 = 1.0 \times 10^{-6}$ T. 
Comparison with data from Fig. 9 of Cattaert et al. \cite{Cattaert2007} shown as dashed lines.
}
\label{benchmark_Cattaert2007Fig9}
\end{figure}

We also benchmark the BO-PBK solver using electron whistlers in moderately magnetized plasmas ($|\omega_{ce}|/\omega_{pe} = 0.5$) from Cattaert et al. \cite{Cattaert2007}. Figure \ref{benchmark_Cattaert2007Fig9} shows the dispersion relation and damping/growth rates under different electron temperature anisotropies.

\subsection*{Case 3: Firehose instability}
\begin{figure}[t]
\centering
\includegraphics[scale=0.5]{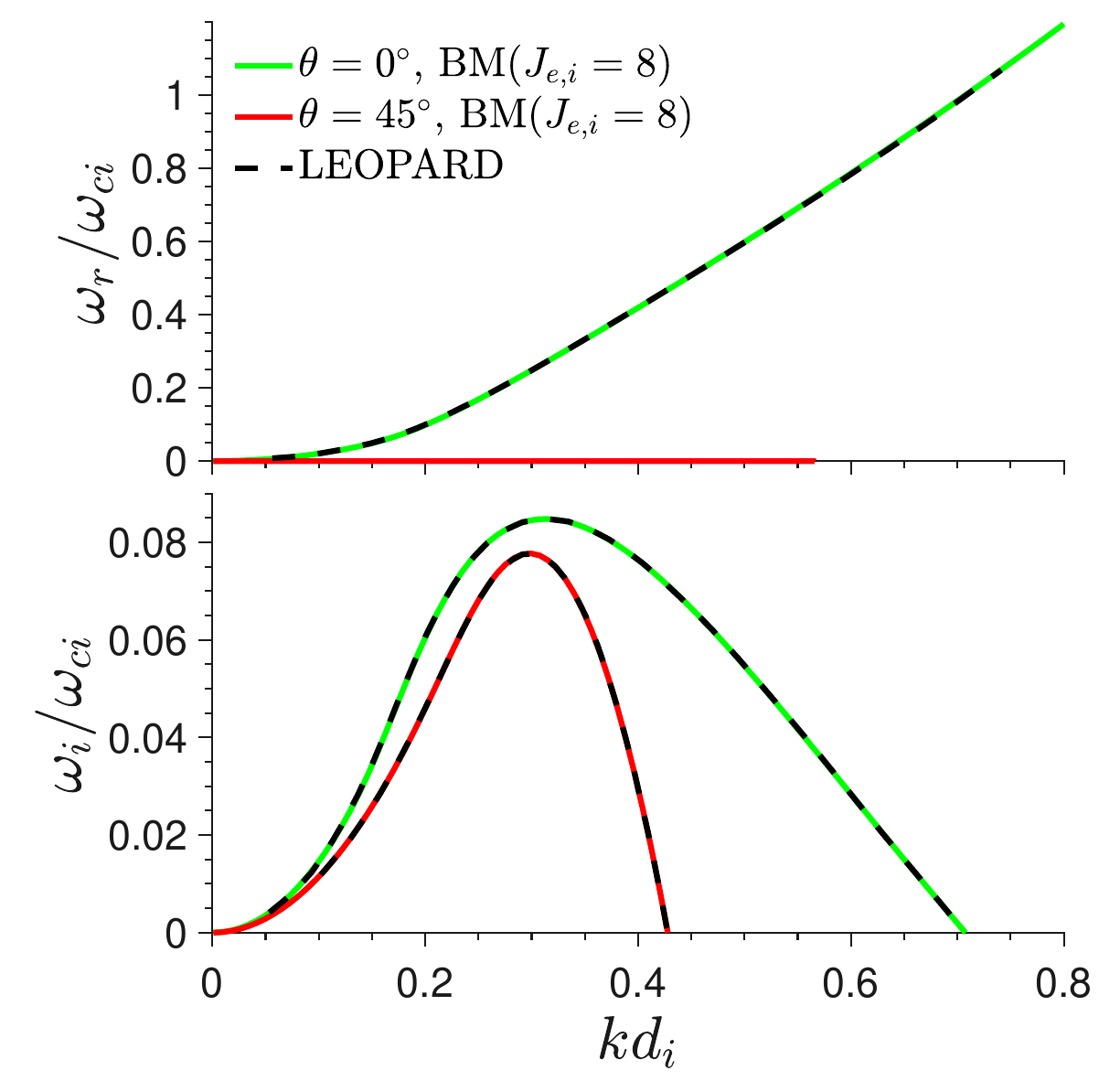}
\caption{
Benchmark of PFHI ($\theta = 0^\circ$) and OFHI ($\theta = 45^\circ$) for a BM distribution ($\beta_{\parallel i} = 4$, $\beta_{\perp i} = 2$, and $\beta_e = 1$). Top: real frequency $\omega_r$;  bottom: growth rate $\omega_i$. The black dashed lines represent the LEOPARD data from Fig. 1 in Ref. \cite{Astfalk2017}.
}
\label{benchmark_Astfalk2017Fig1}
\end{figure}

\begin{figure}[t]
\centering
\includegraphics[scale=0.45]{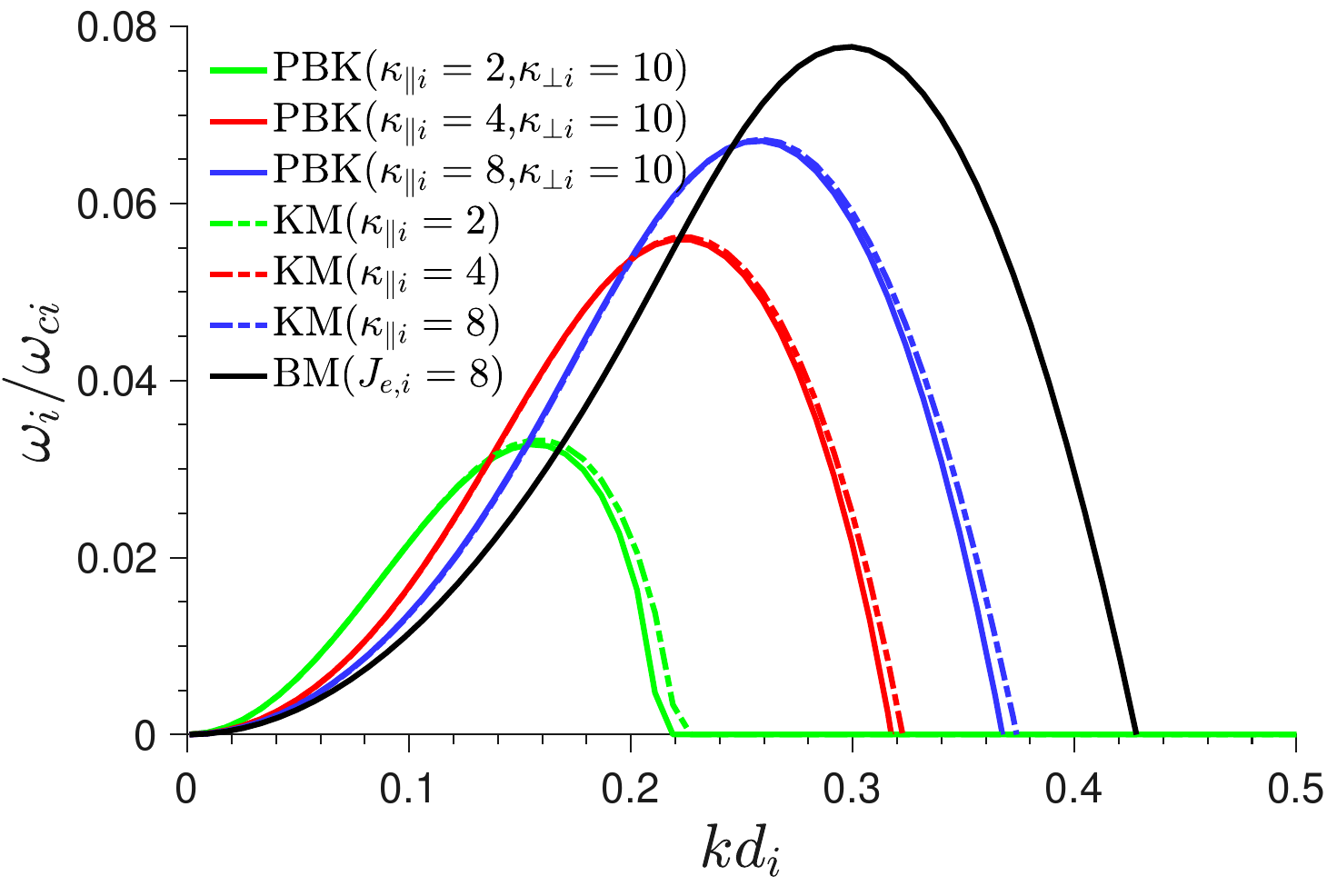}
\caption{
Comparison of OFHI ($\theta = 45^\circ$) for ion distributions: BM (black solid line), KM ($\kappa_{\parallel i}=2,4,8$, $\kappa_{\perp i}=200$, dashed green, red and blue lines) and PBK($\kappa_{\parallel i}=2,4,8$, $\kappa_{\perp i}=10$, solid green, red and blue lines).
}
\label{benchmark_OFHI}
\end{figure}

\begin{figure}[t]
\centering
\includegraphics[scale=0.43]{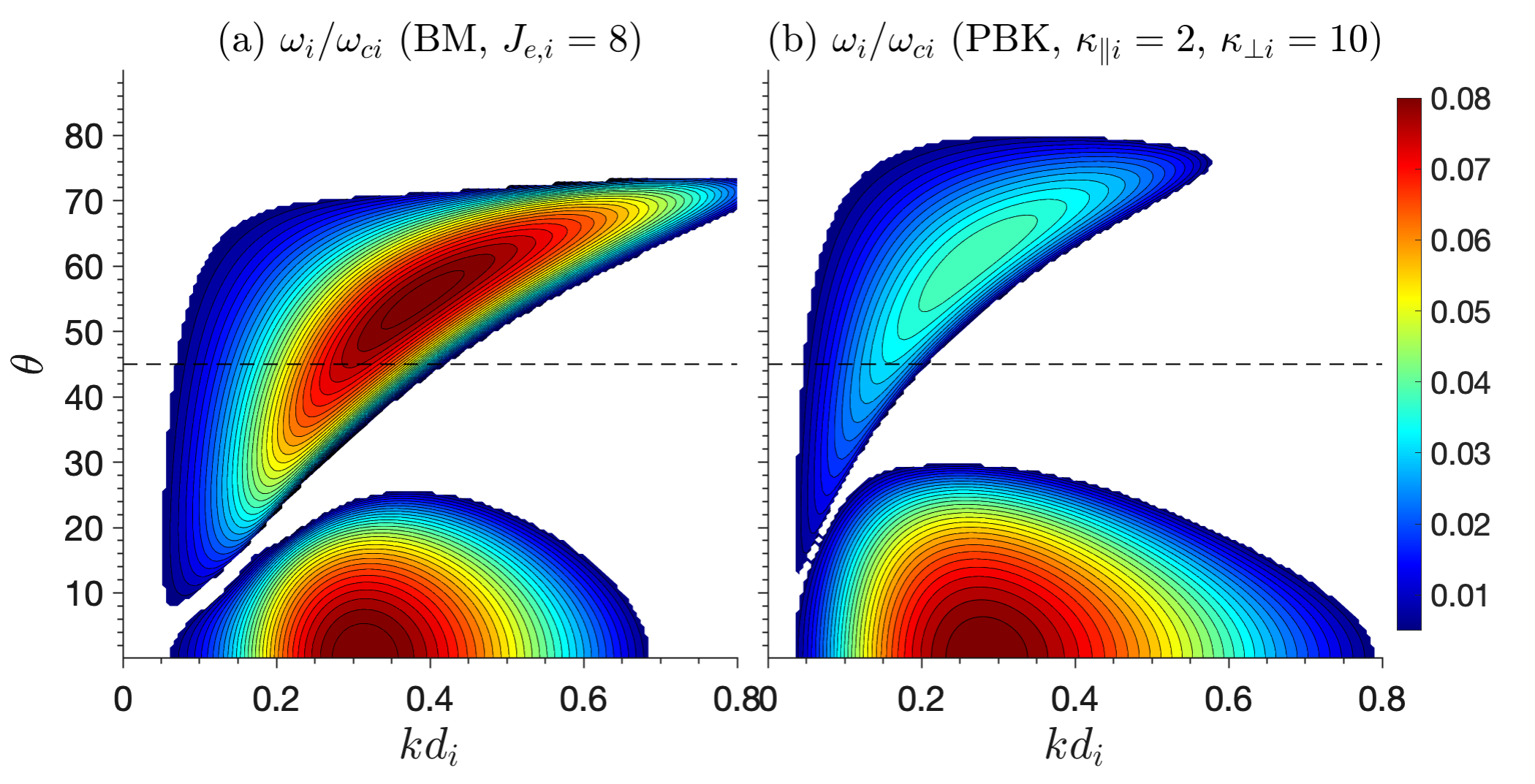}
\caption{
Comparison of the PFHI and OFHI for ions with (a) BM and (b) PBK distributions at various propagation angles.
}
\label{Astfalk2017Fig1_contour}
\end{figure}

We benchmark the BO-PBK code against the parallel (PFHI) and oblique (OFHI) firehose instability cases from the LEOPARD code \cite{Astfalk2015}. The thermal plasma consists of electrons and ions with following parameters\cite{Astfalk2017, bai2025}: electrons have an isotropic BM distribution with $\beta_e = 1$, while ions follow an anisotropic distribution (BM or KM) with $\beta_{\parallel i} = 4$ and $\beta_{\perp i} = 2$. Figure \ref{benchmark_Astfalk2017Fig1} compares BO-PBK results with Ref. \cite{Astfalk2017}.
The ion inertial length is $d_i = v_A/\omega_{ci}$, where the Alfvén velocity is $v_A = B_0/\sqrt{\mu_0 m_i n_i}$. Figure \ref{benchmark_OFHI} shows OFHI ($\theta = 45^\circ$) computed by the BO-PBK solver under the same parameters for different ion distributions: BM, KM($\kappa_{\parallel i}=2,4,8$, $\kappa_{\perp i}=200$), and PBK($\kappa_{\parallel i}=2,4,8$, $\kappa_{\perp i}=10$).
Figure \ref{Astfalk2017Fig1_contour} presents contour plots of the PFHI and OFHI for ions with BM (a) and PBK (b) distributions at various propagation angles.

\subsection*{Case 4: Ion cyclotron instability}
\begin{figure}[t]
\centering
\includegraphics[scale=0.6]{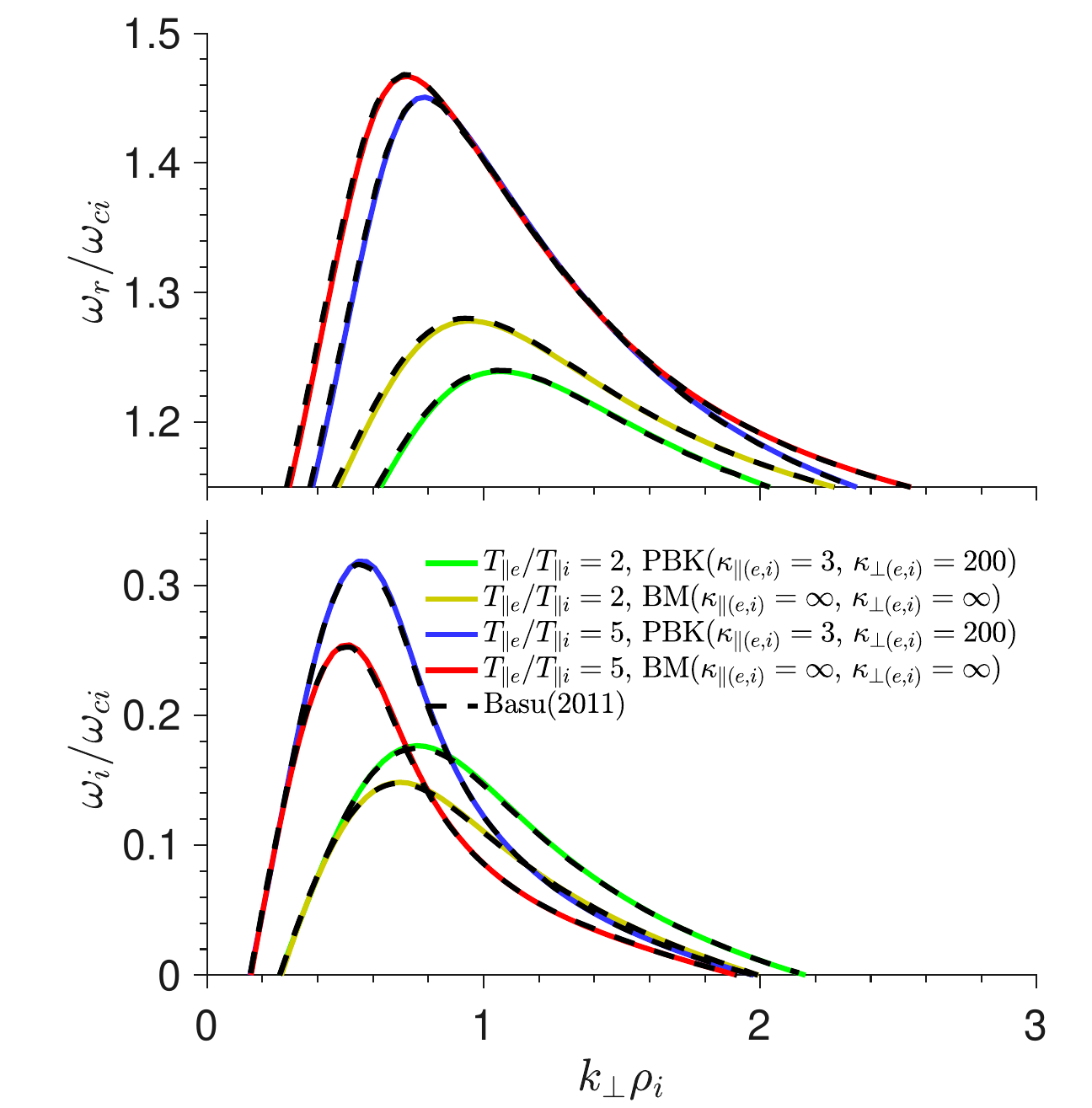}
\caption{
$\omega_r/\omega_{ci}$ vs. $k_{\perp} \rho_i$ (top) and $\omega_i/\omega_{ci}$ vs. $k_{\perp} \rho_i$ (bottom) for $\omega_{pe}/|\omega_{ce}|=1/15$, $u_{e0}/(\tilde{\theta}_{\parallel e}/\sqrt{2}) = 1$,  $k_{\parallel}/(\sqrt{2}\rho_i)=0.08$, $T_{\parallel i}/T_{\perp i}=1$, and $T_{\parallel e}/T_{\parallel i}=2,5$. Solid colored curves:  results from the BO-PBK solver for PBK ($\kappa_{\parallel(e,i)}=3$, $\kappa_{\perp(e,i)}=200$) and BM ($\kappa_{\parallel(e,i)}=\infty$, $\kappa_{\perp(e,i)}=\infty$) plasmas. The black dashed lines are data from Fig. 6 of Basu et al. \cite{Basu2011} for comparison.}
\label{benchmark_Basu2011PoP_fig6}
\end{figure}

\begin{figure}[t]
\centering
\includegraphics[scale=0.6]{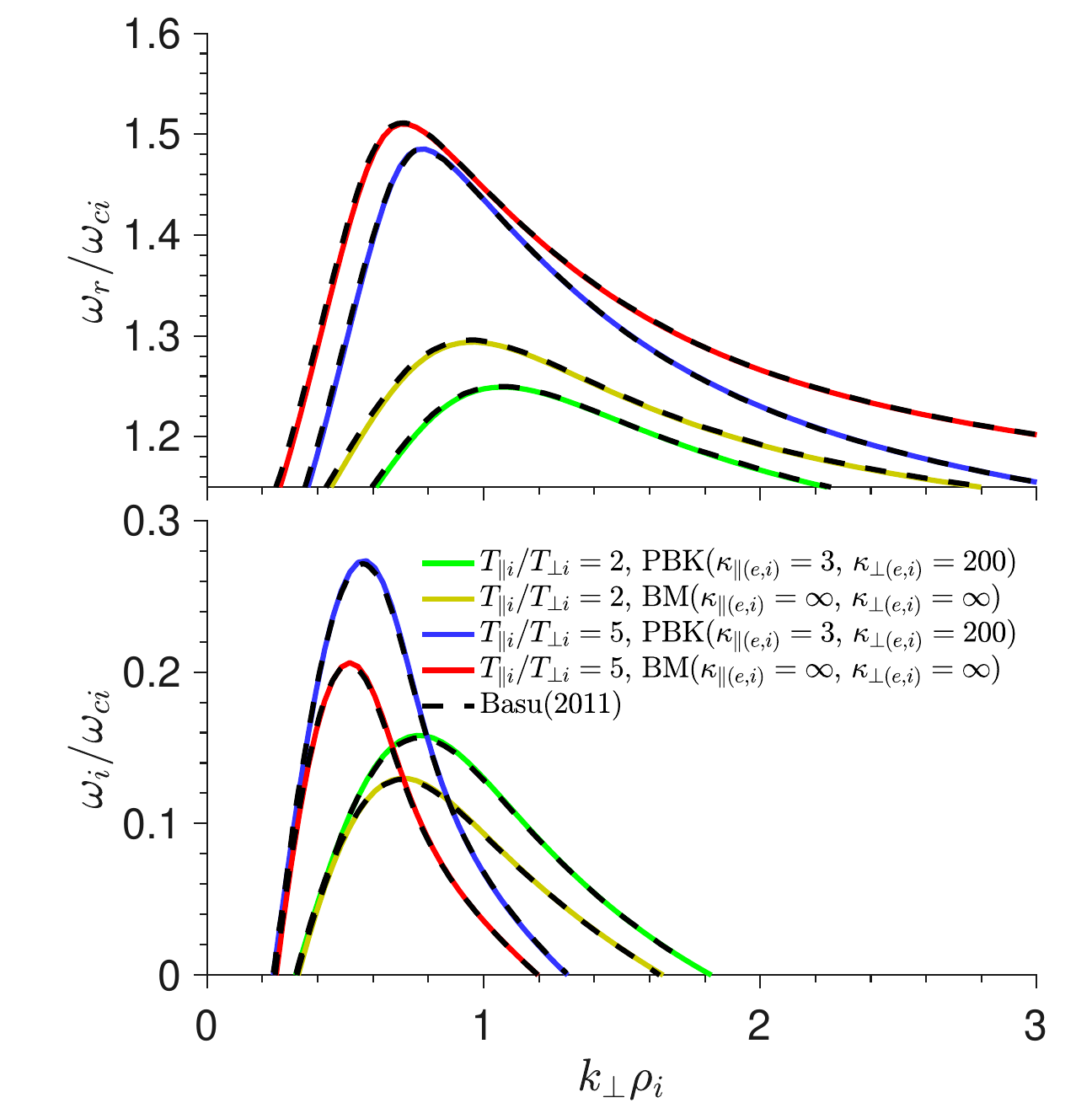}
\caption{
$\omega_r/\omega_{ci}$ vs. $k_{\perp} \rho_i$ (top) and $\omega_i/\omega_{ci}$ vs. $k_{\perp} \rho_i$ (bottom) for $\omega_{pe}/|\omega_{ce}|=1/15$, $u_{e0}/(\tilde{\theta}_{\parallel e}/\sqrt{2}) = 1$,  $k_{\parallel}/(\sqrt{2}\rho_i)=0.08$, $T_{\parallel e}/T_{\parallel i}=1$, and $T_{\parallel i}/T_{\perp i}=2,5$. Solid colored curves:  results from BO-PBK solver for PBK ($\kappa_{\parallel(e,i)}=3$, $\kappa_{\perp(e,i)}=200$) and BM ($\kappa_{\parallel(e,i)}=\infty$, $\kappa_{\perp(e,i)}=\infty$) plasmas. The black dashed lines are data from Fig. 7 of Basu et al. \cite{Basu2011} for comparison.
}
\label{benchmark_Basu2011PoP_fig7}
\end{figure}

\begin{figure}[t]
\centering
\includegraphics[scale=0.5]{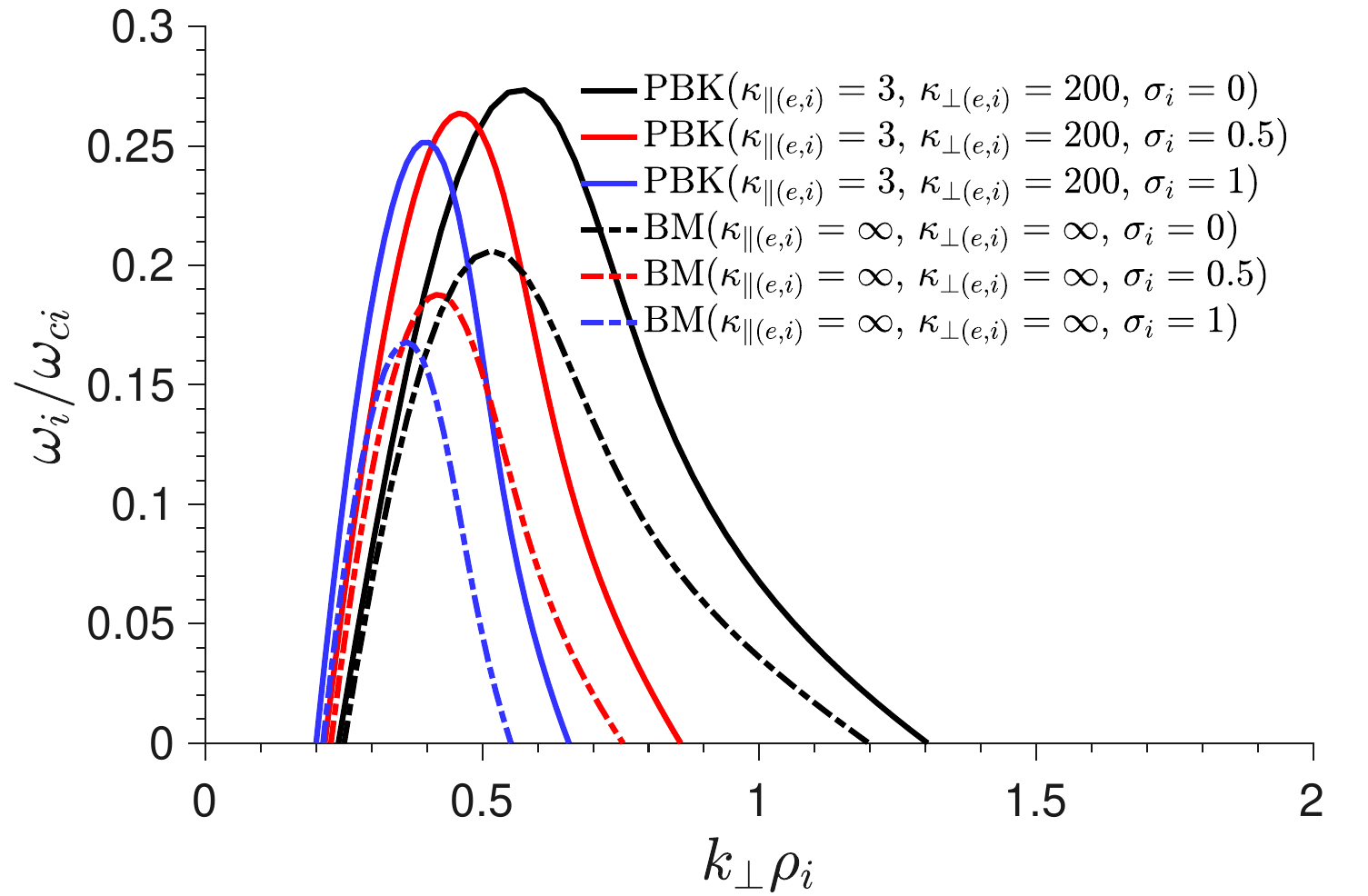}
\caption{$\omega_i/\omega_{ci}$ vs. $k_{\perp} \rho_i$ for parameters $\omega_{pe}/|\omega_{ce}|=1/15$, $u_{e0}/(\tilde{\theta}_{\parallel e}/\sqrt{2}) = 1$,  $k_{\parallel}/(\sqrt{2}\rho_i)=0.08$, $T_{\parallel e}/T_{\parallel i}=1$, and $T_{\parallel i}/T_{\perp i}=5$. 
Solid curves show results for the PBK model ($\kappa_{\parallel(e,i)}=3$, $\kappa_{\perp(e,i)}=200$) and dashed curves for the BM model ($\kappa_{\parallel(e,i)}=\infty$, $\kappa_{\perp(e,i)}=\infty$) both for a loss-cone distribution with $\sigma_{i}=0$ (black), $0.5$ (red), and $1$ (blue).}
\label{benchmark_Basu2011PoP_fig7_sig}
\end{figure}

The current-driven ion cyclotron instability plays a critical role in plasma heating processes in both space and laboratory plasmas. We benchmark the BO-PBK solver against the computational results of Basu \cite{Basu2011}. Figures \ref{benchmark_Basu2011PoP_fig6} and \ref{benchmark_Basu2011PoP_fig7} show the spectral behavior ($\omega_r$ and $\omega_i$ versus $k_{\perp}$ at $k_{\parallel}/(\sqrt{2}\rho_i)=0.08$) of the unstable ion-cyclotron modes for selected $T_{\parallel e}/T_{\perp i}$ and $T_{\parallel i}/T_{\perp i}$ values. Figure \ref{benchmark_Basu2011PoP_fig7_sig} compares the growth rates from the BM and PBK models for different ion loss-cone parameters ($\sigma_{i}=0, 0.5, 1$) at $T_{\parallel i}/T_{\perp i}=5$,  The results demonstrate that the ion loss-cone reduces the growth rate of in this mode.


\begin{figure}[t]
\centering
\includegraphics[scale=0.3]{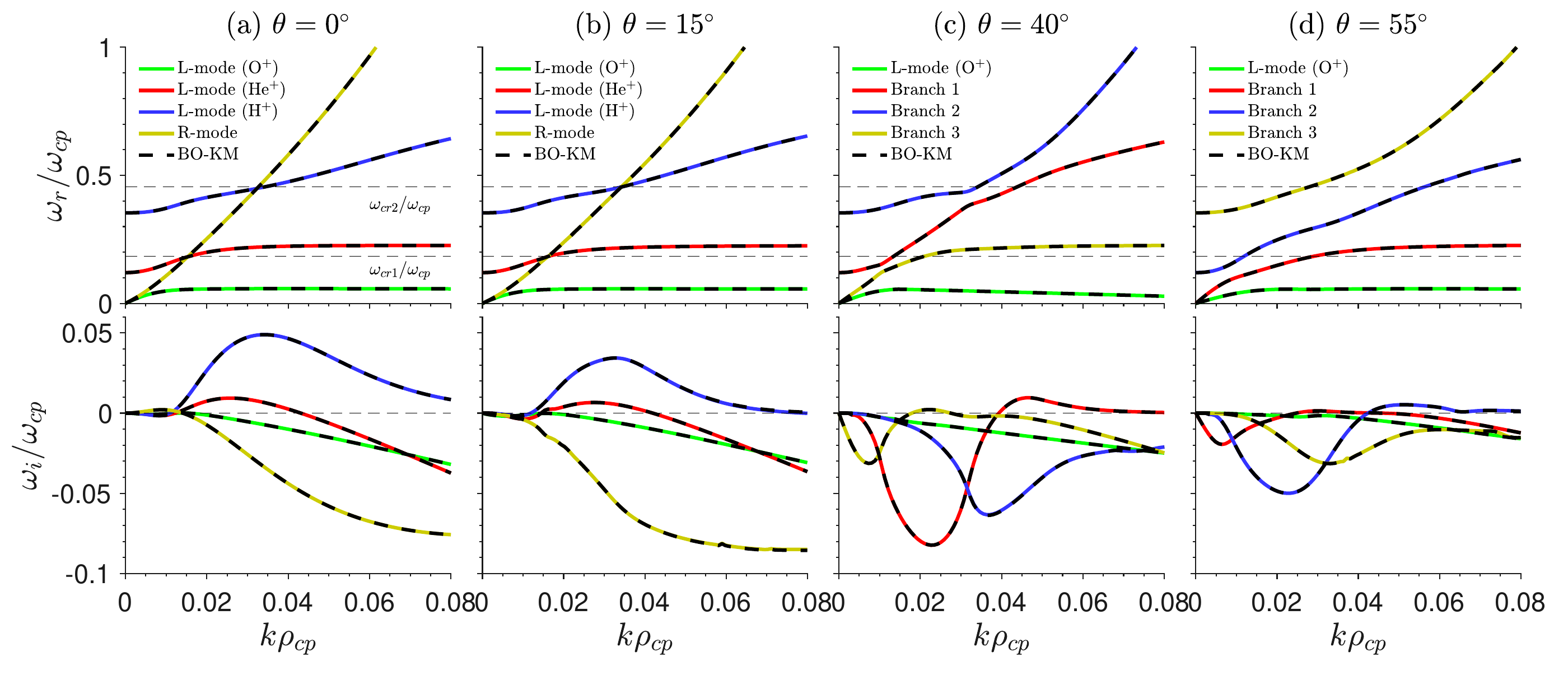}
\caption{$\omega_r/\omega_{cp}$ vs. $k\rho_{cp}$ (top) and $\omega_i/\omega_{cp}$ vs. $k\rho_{cp}$ (bottom) for EMIC waves at propagation angles (a) $\theta = 0^\circ$, (b) $15^\circ$, (c) $40^\circ$, and (d) $55^\circ$. Solid colored curves show BO-PBK solver results, while black dashed lines indicate data from Fig. 13 of Bai et al. \cite{bai2025}. Lower and upper dashed lines in the top panels denote the first and second crossover frequencies, $\omega_{cr1}/\omega_{cp}$ and $\omega_{cr2}/\omega_{cp}$.}
\label{benchmark_Bai2025Fig13}
\end{figure}

To further validate the BO-PBK solver, we examine electromagnetic ion cyclotron (EMIC) waves driven by temperature-anisotropic superthermal protons in a multi-particle plasma composed of cold $\mathrm{H}^+$, $\mathrm{He}^+$, and $\mathrm{O}^+$ ions \cite{Sugiyama2015}. Using plasma parameters from Refs. \cite{Sugiyama2015,bai2025}, we compute the dispersion relations and the growth rates of the $\mathrm{H}^+$-band (dominant) and $\mathrm{He}^+$-band (subdominant) EMIC instabilities for oblique propagation angles $\theta = 0^{\circ}, 15^{\circ}, 40^{\circ}, 55^{\circ}$ (Figs. \ref{benchmark_Bai2025Fig13}(a)-(d)). The BO-PBK solver results (colored solid curves) for superthermal protons with a KM distribution ($\kappa_{\parallel p}=1$ and $\kappa_{\perp p}=200$), show excellent agreement with the BO-KM benchmarks \cite{bai2025} (black dashed lines).
With parameters ($S=4$, $J=8$, and $N=1$) identical to the BO-KM solver \cite{bai2025}, computing all $100$ wave vector points and $60$ angle points takes approximately $76$ minutes of CPU time, which is only about $40\%$ of that required by BO-KM. The efficiency advantage of BO-PBK over BO-KM becomes even more pronounced for large values of $\kappa_{\parallel s}$.

\section{Summary and discussion}
\label{sec5}
This paper presents an enhanced BO-PBK solver for analyzing waves and instabilities in magnetized multi-component plasmas with non-Maxwellian, anisotropic features like power-law tails. The solver models plasmas as temperature-anisotropic conductive media within a unified framework. By linearizing the Vlasov-Maxwell equations with rational dispersion functions ($\mathcal Z_{\kappa_{\parallel s}}$ and $Z_J$), it constructs a concise,  self-consistent linear system, thereby casting the dispersion relation as a standard eigenvalue problem. This enables the solver to handle a wide range of velocity distributions, including anisotropic, loss-cone, drift PBK, KM, BM, and hybrid models.

In BO-PBK solver, adapts to different plasma regimes based on key parameters: the $s$th component approximates a KM plasma for large $\kappa_{\perp s}$, and is treated as BM plasma when $\kappa_{\parallel s}$ exceeds a predefined threshold $\kappa_{\parallel s}^\text{threshold}$. User-defined simulations are configured via the bopbk.in input file. The solver's performance was evaluated against benchmark cases involving waves and instabilities from distributions, including drift, temperature-anisotropic, and multi-component hybrid distributions. The excellent agreement of. the numerical results with established references \cite{Lazar_Poedts2010,Cattaert2007,Sugiyama2015,bai2025} validates the effectiveness and accuracy of the new BO-PBK solver.

In summary, the BO-PBK solver efficiently computes multiple eigenroots of the dispersion relation for obliquely propagating waves in magnetized plasmas featuring high-energy tail distributions. It resolves multiple key plasma wave and instability modes simultaneously, eliminating the need for iterative initial-value searches. The solver is uniformly applicable to hybrid distribution models, such as PBK, KM, and BM, making it well-suited for analyzing dispersion properties and instabilities in both space and laboratory plasmas.
Compared to the BO-KM solver, the present method extends applicability to loss-cone and PBK plasmas and reduces the PBK linear system to three sub-matrices ($\boldsymbol{M}_{x}$, $\boldsymbol{M}_{y}$, and $\boldsymbol{M}_{z}$). Moreover, the proposed transformation method also generalizes to eigenvalue problems arising from dispersion relations in plasmas with arbitrary velocity distributions. Future work will be report related advances.

The source code for this work is available at https://github.com/baiweiphys/BOPBK.

\section*{Acknowledgments}
We are grateful to Jinsong Zhao and Yu Liu for their insightful discussions and valuable suggestions on this research.

\appendix
\section{Proof of the equivalence of linear subsystems}
\label{app1}

From (\ref{xsnlj_matrix_kappa}), we have
\begin{eqnarray}
\label{proof_jx}
\begin{split}
\delta  J_{x}^\mathrm{PBK} =& \frac{b_{x10}}{\omega} \delta E_x  
+ \sum_{sn} \sum_{l=1}^{\kappa_{\parallel s}+1}
\frac{x_{snl1}}{\omega} 
= \frac{b_{x10}}{\omega} \delta E_x 
+ \sum_{sn} \sum_{l=1}^{\kappa_{\parallel s}+1}
\frac{x_{snl2}}{\omega\left(\omega - c_{sn}\right)} 
\\
=& \cdots
= \frac{b_{x10}}{\omega} \delta E_x
+ \sum_{sn} \sum_{l=1}^{\kappa_{\parallel s}+1}
\frac{x_{snll}}{\omega\left(\omega - c_{sn}\right)^{l-1}} 
\\
=& \frac{b_{x10}}{\omega} \delta E_x
+ \sum_{sn}  \sum_{l=1}^{\kappa_{\parallel s}} 
\frac{b_{x33snl,l+1}}{\omega\left(\omega - c_{sn}\right)^{l}}\delta E_z 
\\
&+ \sum_{sn} \sum_{l=1}^{\kappa_{\parallel s}+1}  \left[
\frac{b_{x11snl} \delta E_x}{\omega\left(\omega - c_{sn}\right)^{l}}
+ \frac{b_{x21snl} \delta E_y}{\omega\left(\omega - c_{sn}\right)^{l}}
+ \frac{b_{x31snl} \delta E_z}{\omega\left(\omega - c_{sn}\right)^{l}}
+ \frac{x_{snl,l+1}}{\omega\left(\omega - c_{sn}\right)^{l}} \right]
\\
=& \frac{b_{x10}}{\omega} \delta E_x
+ \sum_{sn} \sum_{l=1}^{\kappa_{\parallel s}} 
\frac{b_{x33snl}}{\omega\left(\omega - c_{sn}\right)^{l-1}}\delta E_z 
\\
&+ \sum_{sn} \sum_{l=1}^{\kappa_{\parallel s}+1} \left[ 
\frac{b_{x11snl} \delta E_x}{\omega\left(\omega - c_{sn}\right)^{l}}
+ \frac{b_{x21snl} \delta E_y}{\omega\left(\omega - c_{sn}\right)^{l}}
+ \frac{b_{x31snl} \delta E_z}{\omega\left(\omega - c_{sn}\right)^{l}} \right]
\\
&+ \sum_{sn} \sum_{l=1}^{\kappa_{\parallel s}+1} \left[ 
\frac{b_{x12snl} \delta E_x}{\omega\left(\omega - c_{sn}\right)^{l+1}}
+ \frac{b_{x22snl} \delta E_y}{\omega\left(\omega - c_{sn}\right)^{l+1}}
+ \frac{b_{x32snl} \delta E_z}{\omega\left(\omega - c_{sn}\right)^{l+1}} \right].
\end{split}
\end{eqnarray}

Similarly, from (\ref{zsnlj_matrix_kappa}), we obtain,
\begin{eqnarray}
\label{proof_jz}
\begin{split}
\delta  J_{z}^\mathrm{PBK} =& i\epsilon_0 \sum_s \omega_{ps}^2 \delta E_z  
+ \sum_{sn} \sum_{l=1}^{\kappa_{\parallel s}+1}
\frac{z_{snl1}}{\omega} 
=  i\epsilon_0 \sum_s \omega_{ps}^2 \delta E_z  
+ \sum_{sn} \sum_{l=1}^{\kappa_{\parallel s + 1}}
\frac{z_{snl2}}{\omega\left(\omega - c_{sn}\right)} 
\\
=& \cdots
=  i\epsilon_0 \sum_s \omega_{ps}^2 \delta E_z  
+ \sum_{sn} \sum_{l=1}^{\kappa_{\parallel s}+1}
\frac{z_{snll}}{\omega\left(\omega - c_{sn}\right)^{l-1}} 
\\
=&  i\epsilon_0 \sum_s \omega_{ps}^2 \delta E_z  
+ \sum_{sn}  \sum_{l=1}^{\kappa_{\parallel s}-1} 
\frac{b_{z34sn,l+2}}{\omega\left(\omega - c_{sn}\right)^{l}}\delta E_z 
\\
&+ \sum_{sn} \sum_{l=1}^{\kappa_{\parallel s}}  \left[
\frac{b_{z13sn,l+1} \delta E_x}{\omega\left(\omega - c_{sn}\right)^{l}}
+ \frac{b_{z23sn,l+1} \delta E_y}{\omega\left(\omega - c_{sn}\right)^{l}}
+ \frac{b_{z33sn,l+1} \delta E_z}{\omega\left(\omega - c_{sn}\right)^{l}} \right]
\\
&+ \sum_{sn} \sum_{l=1}^{\kappa_{\parallel s}+1}  \left[
\frac{b_{z11snl} \delta E_x}{\omega\left(\omega - c_{sn}\right)^{l}}
+ \frac{b_{z21snl} \delta E_y}{\omega\left(\omega - c_{sn}\right)^{l}}
+ \frac{b_{z31snl} \delta E_z}{\omega\left(\omega - c_{sn}\right)^{l}}
+ \frac{z_{snl,l+1}}{\omega\left(\omega - c_{sn}\right)^{l}} \right]
\\
=& i\epsilon_0 \sum_s \omega_{ps}^2 \delta E_z  
+ \sum_{sn} \sum_{l=1}^{\kappa_{\parallel s}-1} 
\frac{b_{z34sn,l+2}}{\omega\left(\omega - c_{sn}\right)^{l-1}}\delta E_z 
\\
&+ \sum_{sn} \sum_{l=1}^{\kappa_{\parallel s}}  \left[
\frac{b_{z13sn,l+1} \delta E_x}{\omega\left(\omega - c_{sn}\right)^{l}}
+ \frac{b_{z23sn,l+1} \delta E_y}{\omega\left(\omega - c_{sn}\right)^{l}}
+ \frac{b_{z33sn,l+1} \delta E_z}{\omega\left(\omega - c_{sn}\right)^{l}} \right]
\\
&+ \sum_{sn} \sum_{l=1}^{\kappa_{\parallel s}+1} \left[ 
\frac{b_{z11snl} \delta E_x}{\omega\left(\omega - c_{sn}\right)^{l}}
+ \frac{b_{z21snl} \delta E_y}{\omega\left(\omega - c_{sn}\right)^{l}}
+ \frac{b_{z31snl} \delta E_z}{\omega\left(\omega - c_{sn}\right)^{l}} \right]
\\
&+ \sum_{sn} \sum_{l=1}^{\kappa_{\parallel s}+1} \left[ 
\frac{b_{z12snl} \delta E_x}{\omega\left(\omega - c_{sn}\right)^{l+1}}
+ \frac{b_{z22snl} \delta E_y}{\omega\left(\omega - c_{sn}\right)^{l+1}}
+ \frac{b_{z32snl} \delta E_z}{\omega\left(\omega - c_{sn}\right)^{l+1}} 
\right].
\end{split}
\end{eqnarray}

\section{Useful formulas}
\label{app2}
(a) Limit formulas
\begin{equation}
\label{Limit_eqs}
\lim_{\kappa \rightarrow \infty} \left(1 + \frac{y^2}{\kappa} \right)^{-\kappa}
= e^{-y^2},
\quad 
\lim_{n \rightarrow \infty}  \frac{\Gamma(n+z)}{\Gamma(n) n^z} = 1.
\end{equation}

(b) Gamma function
\begin{equation}
\label{Gamma_func}
\Gamma(z+1) = z\Gamma(z),
\quad
\Gamma(n) = (n-1)!, 
\quad
\Gamma(n+\frac{1}{2}) = \sqrt{\pi}\frac{(2n)!}{4^n n!}.
\end{equation}

(c) Summation formulas for Bessel function \cite{stix1992waves}
\begin{eqnarray}
\begin{cases}
\sum_{n=-\infty}^{\infty} n J_n^2 = 0, 
\quad
\sum_{n=-\infty}^{\infty} J_n J_n^{\prime} = 0, 
\quad 
\sum_{n=-\infty}^{\infty} J_n^2 = 1,
\\
\sum_{n=-\infty}^{\infty} \left(J_n^{\prime}\right)^2 = \frac{1}{2}, 
\quad
\sum_{n=-\infty}^{\infty} n^2 J_n^2(z) =  \frac{1}{2} z^2, 
\quad
\sum_{n=-\infty}^{\infty} n J_n J_n^{\prime} = 0,
\\
\sum_{n=-\infty}^{\infty} I_n(z) = \sum_{n=-\infty}^{\infty} I^{\prime}_n(z)  = \exp(z),
\quad
\sum_{n=-\infty}^{\infty} n^2 I_n(z) =  z\exp(z),
\\
\sum_{n=-\infty}^{\infty} n I_n = \sum_{n=-\infty}^{\infty} n I^{\prime}_n = 0.
\end{cases}
\end{eqnarray}

(d) Gamma function Integrals
\begin{subequations}
\begin{align}
&\int_{0}^{\infty} \frac{x^{\mu-1}}{\left(p+q x^2\right)^{\nu}}dx
= \frac{\Gamma(\mu/2) \Gamma(\nu-\mu/2)}{2 \Gamma(\nu)}
\frac{1}{q^{\mu/2}} \frac{1}{p^{(\nu-\mu/2)}},
\\
&\int_{0}^{\infty} x^{\mu-1} e^{-\nu x^2} dx
= \frac{\Gamma(\mu/2) }{2 \nu^{\mu/2}}.
\end{align}
\end{subequations}

(e) Integrals involving modified Bessel functions
\begin{eqnarray}
\begin{cases}
\int_{0}^{\infty} x J_{n}^2(x) \exp\left(-\frac{x^2}{2\lambda} \right) d x
= \lambda \Lambda_n,
\\
\int_{0}^{\infty} x^2 J_{n}(x) J_{n}^{\prime}(x) \exp\left(-\frac{x^2}{2\lambda} \right) d x
= \lambda^2 \Lambda_n^{\prime},
\\
\int_{0}^{\infty} x^3 J_{n}^{2\prime}(x) \exp\left(-\frac{x^2}{2\lambda} \right) d x
= n^2\lambda \Lambda_n-2\lambda^3 \Lambda_n^{\prime},
\end{cases}
\end{eqnarray}
where $\lambda >0$; $n=0,\pm1,\pm2,\cdots$. Here, $\Lambda_n$ and its derivative are given by
\begin{equation}
\Lambda_n = I_n(\lambda) e^{-\lambda},
\quad
\Lambda_n^{\prime} = I_n^{\prime}(\lambda) e^{-\lambda} - I_n(\lambda) e^{-\lambda}.
\end{equation}
where $I_n$ is the modified Bessel function of the first kind of order $n$, whose derivative obeys
\begin{eqnarray}
I_n^{\prime}(\lambda) = \frac{1}{2}\left[I_{n-1}(\lambda) + I_{n+1}(\lambda) \right].
\end{eqnarray}






\bibliographystyle{ans}
\bibliography{./myRefs}



\end{document}